\documentstyle[epsfig,aps,floats,eqsecnum,preprint]{revtex}

 \def\tdot#1{{\buildrel{\ldots}\over{#1}}}
\newcommand{\vrsmall}{\vrule width 0pt height 18pt depth 15pt}
\newcommand{\vrbig}{\vrule width 0pt height 27pt depth 15pt}

\newcommand{\pa}{\partial} 
\newcommand{\co}{\nabla}
\newcommand{\vphi}{\varphi} 
\newcommand{\sigmap}{\sigma^\prime}
\newcommand{\taup}{\tau^\prime}
\newcommand{\beq}{\begin{equation}} 
\newcommand{\eeq}{\end{equation}}
\newcommand{\bea}{\begin{eqnarray}} 
\newcommand{\eea}{\end{eqnarray}}
\newcommand{\beam}{\begin{mathletters}} 
\newcommand{\eeam}{\end{mathletters}}

\begin{document}
\draft 
\title{\Large\bf On the gravitational, dilatonic and axionic radiative
damping of cosmic strings} \author{Alessandra Buonanno$^1$ and
Thibault Damour$^{1,2}$}

\address{$^1$ {\it Institut des Hautes Etudes Scientifiques, 91440
Bures-sur-Yvette, France} \\ {$^2$ {\it DARC, CNRS-Observatoire de
Paris, 92195 Meudon, France}}}
\vskip 1.5truecm \maketitle
\begin{abstract}
We study the radiation reaction on cosmic strings due to the emission
of dilatonic, gravitational and axionic waves. After verifying the 
(on average) conservative nature of the time-symmetric 
self-interactions, we concentrate on the finite radiation damping 
force associated with the half-retarded minus half-advanced ``reactive'' 
fields. We revisit a recent proposal of using a ``local back reaction 
approximation'' for the reactive fields. Using {\it dimensional 
continuation} as convenient technical tool,  
we find, contrary to previous claims, 
that this proposal leads to {\it antidamping}\/ in the case of the axionic 
field, and to {\it zero} (integrated) {\it damping} in the case of the 
gravitational field. One gets normal {\it positive damping} only 
in the case of the dilatonic field.
We propose to use a suitably modified version of the local dilatonic radiation 
reaction as a substitute for the exact (non-local) gravitational radiation 
reaction. The incorporation of such a local approximation to gravitational 
radiation reaction should allow one to complete, in a computationally 
non-intensive way, string network simulations and to give better estimates of the 
amount and spectrum of gravitational radiation emitted by a cosmologically 
evolving network of massive strings.
\end{abstract}
\vskip 0.5truecm
\pacs{PACS: 98.80.Cq \hskip 2cm IHES/P/98/06 \hskip 2cm gr-qc/9801105}

\section{Introduction}
\label{sec1}
Cosmic strings are predicted, within a wide class of elementary particle 
models, to form at phase transitions in the early universe 
\cite{HindmarshKibble95}, \cite{VilenkinShellard94}. The creation of a network 
of cosmic strings can have important astrophysical consequence, notably for 
the formation of structure in the universe \cite{Zeldowich80}, 
\cite{Vilenkin81a}. A network of cosmic strings might also be a copious source 
of the various fields or quanta to which they are coupled. Oscillating loops 
of 
cosmic string can generate observationally significant stochastic backgrounds 
of: gravitational waves \cite{Vilenkin1981d}, massless Goldstone bosons 
\cite{Davis1985a}, light axions \cite{Davis1986}, \cite{DavisShellard1989c}, 
or 
light dilatons \cite{DamourVilenkin97}. The amount of radiation emitted by 
cosmic strings depend: (i) on the nature of the considered field; (ii) on the 
coupling parameter of this field to the string; (iii) on the dynamics of 
individual strings; and (iv) on the distribution function and cosmological 
evolution of the string network. It is important to note that the latter 
network distribution function in turn depends on the radiation properties of 
strings. Indeed, numerical simulations suggest that the characteristic size of 
the loops chopped off long strings at the epoch $t$ will be on order of the 
smallest structures on the long strings, which is itself arguably determined 
by 
radiative back reaction \cite{BennettBouchet88}, \cite{QuashnockSpergel90}. 
For 
instance, if one considers grand unified theory (GUT) scale strings, with 
tension $\mu \sim \Lambda_{\rm GUT}^2$, gravitational radiation (possibly 
together with dilaton radiation which has a comparable magnitude 
\cite{DamourVilenkin97}) will be the dominant radiative mechanism, and will be 
characterized by the coupling parameter $G\mu \sim (\Lambda_{\rm GUT} / m_{\rm 
Planck})^2 \sim 10^{-6}$. It is then natural to expect that the same 
dimensionless parameter $G\mu$ will control the radiative decay of the small 
scale structure (crinkles and kinks) on the horizon-sized strings, thereby 
determining also the characteristic size relative to the horizon of the small 
loops produced by the intersections of long strings: $\ell_{\rm loops} \equiv 
\widehat{\alpha} \, ct$, with $\widehat{\alpha} \sim \Gamma_{\rm kink} \, 
G\mu$, $\Gamma_{\rm kink}$ being some dimensionless measure of the 
network-averaged radiation efficiency of kinky strings \cite{BennettBouchet88}, 
\cite{QuashnockSpergel90}, \cite{AllenCaldwell91}, 
\cite{AustenCopelandKibble93}. If one considers ``global'' strings, i.e. 
strings formed when a global symmetry is broken at a mass scale $f_a$, 
emission 
of the Goldstone boson associated to this symmetry breaking will be the 
dominant radiation damping mechanism and will be characterized by the 
dimensionless parameter $f_a^2 / \mu_{\rm effective} \sim (\log 
(L/\delta))^{-1} \sim 10^{-2}$, where the effective tension 
$\mu_{\rm effective}$ is  renormalized by a large logarithm (see, e.g., 
\cite{VilenkinShellard94}).

Present numerical simulations of string networks do not take into account the 
effect of radiative damping on the actual string motion. The above mentioned 
argument concluding in the case of GUT strings to the link $\widehat{\alpha} 
\sim \Gamma_{\rm kink} \, G\mu$ between the loop size and radiative effects 
has 
been justified by Quashnock and Spergel \cite{QuashnockSpergel90} who studied 
the gravitational back reaction of a sample of cosmic string loops. However, 
their ``exact'', non-local approach to gravitational back reaction is 
numerically so demanding that there is little prospect to implementing it in 
full string network simulations. This lack of consideration of the dynamical 
effects of radiative damping is a major deficiency of string network 
simulations which leaves unanswered crucial questions such as: Is the string 
distribution function attracted to a solution which ``scales'' with the 
horizon 
size down to the smallest structures? and What is the precise amount and 
spectrum of the gravitational (or axionic, in the case of global strings) 
radiation emitted by the combined distribution of small loops and long 
strings?
 
Recently, Battye and Shellard \cite{BS95}, \cite{BS96} proposed a new, 
computationally much less intensive, approach to the radiative back reaction 
of 
(global) strings. They proposed a ``local back reaction approximation'' based 
on an analogy with the well-known Abraham-Lorentz-Dirac result for a 
self-interacting electron. Their approach assumes that the dominant 
contribution to the back reaction force density at a certain string point 
comes 
from string segments in the immediate vicinity of that point. They have 
endeavored to justify their approach by combining analytical results 
(concerning approximate expressions of the local, axionic radiative damping 
force) and numerical simulations (comparison between the effect of their local 
back reaction and a direct field-theory evolution of some global string 
solutions).

In this paper, we revisit the problem of the back reaction of cosmic strings 
associated to the emission of gravitational, dilatonic and axionic fields, 
with 
particular emphasis on the ``local back reaction approximation'' of Battye and 
Shellard. Throughout this paper, we limit our scope to the self-interaction 
of Nambu strings, in absence of any non-trivial external fields. This problem 
can be (formally) treated by a standard perturbative approach, i.e. by 
expanding all quantities in powers of the gravitational\footnote{
Because of our ``gravitational normalization'' of the kinetic 
terms, see Eq.(\ref{sb}), the couplings of all the three considered fields 
are proportional to $G$.} coupling constant $G$. We work only to first-order 
in $G$. To this order, we first verify the fact (well-known to hold for 
self-interacting, electrically charged, point particles \cite{Dirac38}) 
that the time-symmetric part of the self-interaction (i.e. the part mediated by 
the half-retarded {\it plus} half-advanced Green function) is, 
on the average, conservative, i.e. that it does not 
(after integration) drain energy-momentum out of the string. 
As we are interested in radiation damping, this allows us to concentrate 
on the time-odd part of the self-interaction, mediated by the half-retarded 
{\it minus} half-advanced Green function. This ``reactive'' part of the self-interaction 
is (as in the case of point charges) {\it finite}. [By contrast, the time-symmetric 
self-interaction is (formally) ultraviolet divergent. This divergence is not of concern for us here because, as shown in Refs.~\cite{CB}, \cite{BDletter} and further discussed 
below, its infinite part is renormalizable, and, as said above, its finite part does not 
globally contribute to damping.] Contrary to the case of point charges, the reactive 
part is non-local, being given by an integral over the string. Following Battye and 
Shellard \cite{BS95}, \cite{BS96} we study the ``local approximation'' to this reaction 
effect. We find very convenient for this study to use the technique of {\it dimensional 
continuation} (well-known in quantum field theory). 

In the case of the axionic self-field, we find that the axionic 
reaction force defined by the 
``local back reaction approximation'' of Ref.~\cite{BS95} 
leads to {\it antidamping}\/ rather than damping, as claimed in Refs. \cite{BS95}, 
\cite{BS96}. We also investigate below the 
corresponding local approximations to gravitational and dilatonic self-forces and find {\it 
zero damping} in the gravitational case, and a normal, {\it positive 
damping} for the dilatonic case. The physical origin of these paradoxical results 
is explained below (Sec.~\ref{subsec4.5}) by tracing them to the modification 
of the field propagator implicitly entailed by the use of the local back 
reaction approximation. We show that, in the case of {\it gauge fields}, 
this modification messes up the very delicate sign compensations 
which ensure the positivity of the energy carried away by gauge fields. 
Thereby, one of the main results of the present work is to prove the 
untenability of applying a straightforward local back reaction approximation 
to {\it gauge fields} such as gravitational and axionic fields. 
However, this untenability does not necessarily apply to the case of non-gauge fields.
Indeed, our work proves that the application of this local approximation 
to the dilatonic field (which is not a gauge field) leads to the 
correct sign for damping effects. In this non-gauge field case, the argument 
(of Sec.~\ref{subsec4.5}) which showed the dangers of approximating the 
field propagator for gauge fields, looses its strength. This leaves 
therefore open the question of whether the ``local approximation''
to dilatonic back reaction might define (despite its shortcomings discussed below) 
a phenomenologically acceptable approximation to the exact, non-local 
self-force. In this direction, we give several arguments, and strengthen them 
by some explicit numerical calculations, toward showing that
the meaningful (positive-damping) {\it dilatonic} local back reaction force can be 
used, after some modification, as a convenient effective {\it substitute}\/ for the 
exact (non-local) {\it gravitational} back reaction force.

This phenomenological proposal is somewhat of an expedient because it rests 
on an ``approximation''  whose validity domain is severely limited. 
However, pending the discovery of a better local proposal, 
we think that the incorporation of our proposed local reaction force (\ref{for}) 
should allow one to complete, in a computationally 
non-intensive way, string network simulations and to give better estimates 
of the amount and spectrum of gravitational 
radiation emitted by a cosmologically evolving network of massive strings.

In the next section, we present our formalism for treating self-interactions 
of 
strings. We describe in Section \ref{sec3} our results for the renormalizable, 
divergent self-action terms, and, in Section \ref{sec4}, our results for the finite 
contributions to the ``local'' reaction force. In Section 
\ref{sec5} we indicate 
how the local dilatonic damping force could be used in full-scale network 
simulations to simulate the dynamical effects of gravitational radiation.
Section \ref{sec6} contains our conclusions. Some 
technical details are relegated to the Appendix.

As signs will play a crucial role below, let us emphasize that we use the 
``mostly positive'' signature $(-,+,+,+)$ for the space-time metric $g_{\mu 
\nu}$ ($\mu , \nu = 0,1,2,3$), and the corresponding $(-,+)$ signature for 
the worldsheet metric $\gamma_{ab}$ ($a,b = 0,1$ being worldsheet indices).

\section{Cosmic strings interacting with gravitational, dilatonic and axionic 
fields}
\label{sec2}

We consider a closed Nambu string $z^{\mu} (\sigma^a)$ (with $\sigma^0 = \tau$, 
$\sigma^1 = \sigma$, $0 \leq \sigma < L$) interacting with its own gravitational 
$g_{\mu \nu} (x^{\lambda}) \equiv \eta_{\mu \nu} + h_{\mu \nu} (x^{\lambda})$, 
dilatonic $\varphi (x)$, and axionic (Kalb-Ramond) $B_{\mu \nu} (x)$ fields. 
The action for the string coupled to $g_{\mu \nu}$, $\varphi$ and $B_{\mu 
\nu}$ 
reads
\beq
\label{sa}
S_s = - \int \mu (\varphi) \, d A - \frac{\lambda}{2} \int B_{\mu \nu} \, 
dz^{\mu} \wedge dz^{\nu} \, .
\eeq
Here $dA = \sqrt{\gamma} \, d^2 \, \sigma$ (with $\gamma \equiv -\det 
\gamma_{ab}$; $\gamma_{ab} \equiv g_{\mu \nu} (z) \, \partial_a \, z^{\mu} \, 
\partial_b \, z^{\nu}$ denoting the metric induced on the worldsheet) is the 
string area element and the dilaton dependence of the string tension $\mu$ can 
be taken to be  exponential
\beq
\label{n1}
\mu (\varphi) = \mu \, e^{2\alpha \varphi} \, .
\eeq
At the linearized approximation where 
we shall work the form (\ref{n1}) is equivalent 
to a linear coupling $\mu (\varphi) \simeq \mu (1+2 \alpha \varphi)$. The 
dimensionless parameter $\alpha$ measures the strength of the coupling of 
$\varphi$ to cosmic strings (our notation agrees with the tensor-scalar 
notation of Ref. \cite{DEF92}), while the coupling strength of the axion field 
is measured by the parameter $\lambda$ with dimension $({\rm mass})^2$. Due to 
our ``gravitational normalization'' of the kinetic term of $B_{\mu \nu}$, the 
link between $\lambda$ and the mass scale $f_a$ used in Refs. \cite{BS95}, 
\cite{BS96} is $ 2G \, \lambda^2 = \pi f_a^2$.

The action for the fields is
\beq 
\label{sb}
S_f = \frac{1}{16 \pi G}\, \int d^4 x \sqrt{g}\,\left [
{\cal R} -2\,\co^\mu \vphi\,\co_\mu \vphi -\frac{1}{12}\,e^{-4\alpha\vphi}\,
H_{\mu \nu \rho}\, 
H^{\mu \nu \rho}\,\right ] \,,
\eeq
where $H_{\mu \nu \rho} = \partial_{\mu} \, B_{\nu \rho} + \partial_{\nu} \, 
B_{\rho \mu} + \partial_{\rho} \, B_{\mu \nu}$, $g \equiv -\det \, (g_{\mu 
\nu})$, and where we use the curvature conventions 
${\cal R}^\mu_{\,\,\,\,\nu \rho \sigma} = \pa_\rho \Gamma^\mu_{\nu\sigma}
- \dots\,,\,\,{\cal R}_{\mu \nu} = {\cal R}^\rho_{\,\,\,\,\mu \rho \nu}$.
With this notation, 
a tree-level coupled fundamental string (of string theory) has $\alpha = 1$  
(in 4 dimensions) and  $\lambda = \mu$. 

Everywhere in this paper, we shall assume the absence of external fields. 
More precisely, the background values of the fields we consider are 
$g_{\mu \nu}^{0} = \eta_{\mu \nu}$, $\vphi^0 = 0$,  $B_{\mu \nu}^0 = 0$. 
Our results are derived only for this case, by using (formal) perturbation 
theory around these trivial backgrounds. It is however understood, as usual, 
that one can later (e.g. for cosmological applications) reintroduce 
a coupling to external fields, varying on a scale much larger than the 
size of the string, by suitably covariantizing the final, trivial-background 
results derived here. Such an approximate treatment should be sufficient
for the cosmological applications we have in mind. On the other hand, the methods 
used here are not appropriate for treating the general case of a string 
interacting with external gravitational and dilatonic fields of arbitrary 
strength and spacetime variability. To treat such a case, one would 
need a more general formalism, such as that of Ref.~\cite{BattyeCarter}. 
Note, however, that the straightforward, non-explicitly covariant, perturbation 
approach to radiation damping effects used here is the string analog
\footnote{This analog is technically simpler because radiation damping appears 
at linear order for strings (which have a non-trivial, accelerated motion 
at zeroth order), while it is a non-linear phenomenon in gravitationally 
bound systems.} of all the standard work done on the gravitational 
radiation damping of binary 
systems (see, e.g., \cite{Damour87} for a review).

In other words, our aim in this paper is to derive, consistently at the 
first order in the basic coupling constant $G$ (see Eq.~(\ref{sb})), both the 
fields generated by the string, 
\bea
h_{\mu \nu}(x) \equiv g_{\mu \nu}(x) - \eta_{\mu \nu} &=& 
G\,h_{\mu \nu}^1(x) + {\cal O}(G^2)\,, \quad \quad 
\vphi(x)=  0 + G\,\vphi^1(x) + {\cal O}(G^2)\,, \nonumber \\ 
&& B_{\mu \nu}(x)=  0 + G\,B_{\mu \nu}^1(x) + {\cal O}(G^2)\,,
\label{n2.1}
\eea
and the (non-covariant) explicit form of the string equations of motion, 
written in a specified~\footnote{
We shall use the conformal gauge associated 
to the metric $g_{\mu \nu}(x) = \eta_{\mu \nu} + G\,h_{\mu \nu}^1 + 
{\cal O}(G^2)$.} (class of) worldsheet gauge(s), 
\beq
\label{n2.2}
\mu\,\eta_{\mu \nu}\,(\ddot{z}^\nu - z^{\prime \prime \nu}) = 
G\,{\cal F}_\mu^1 + {\cal O}(G^2)\,.
\eeq
The string action (\ref{sa}) 
can be written (using the Polyakov form) as
\beq
S_{\rm s} = -\frac{\mu}{2}\, \int d^2 \sigma\,e^{2\alpha\,\vphi}\,
\sqrt{\widehat{\gamma}}\,\widehat{\gamma}^{a b}\,
\pa_a z^\mu\,\pa_b z^\nu\,g_{\mu \nu} 
-\frac{\lambda}{2}\, \int d^2 \sigma 
\epsilon^{a b}\, \pa_a z^\mu\,\pa_b z^\nu\,B_{\mu \nu}\,,
\eeq
where the worldsheet metric $\widehat{\gamma}_{ab}$ must be independently 
varied and where $\epsilon^{01} = -1$, $\epsilon^{10} = 1$. The equation of 
motion of $\widehat{\gamma}_{ab}$ is the constraint that it be conformal to 
the 
induced metric $\gamma_{ab} = g_{\mu \nu} (z) \, \partial_a \, z^{\mu} \, 
\partial_b \, z^{\nu}$. In the following, we shall often use the conformal 
gauge $\sqrt{\widehat{\gamma}} \, \widehat{\gamma}^{ab} = \sqrt{\gamma} \, 
\gamma^{ab} = \eta^{ab}$ (where $\eta^{00} = -1$, $\eta^{11} = +1$), i.e. we 
shall choose the $(\tau , \sigma)$ parametrization of the worldsheet so that
\beq
\label{con}
\dot{z}^\mu\,\dot{z}^\nu\,g_{\mu \nu} + 
z^{\prime\,\mu}\,z^{\prime\,\nu}\,g_{\mu \nu}=0 \,, \quad \quad 
\dot{z}^\mu\,z^{\prime\,\nu}\,g_{\mu \nu}=0\,.
\eeq
Here $\dot z \equiv \partial_0 \, z \equiv \partial \, z / \partial \, \tau$ 
and $z' \equiv \partial_1 \, z \equiv \partial \, z / \partial \, \sigma$. 
Note 
also the expression, in this gauge, of the worldsheet volume density
\beq
\sqrt{\gamma} = g_{\mu \nu} \, z^{\prime \,\mu} \, z^{\prime \,\nu} = 
-g_{\mu \nu} \, {\dot 
z}^{\mu} \, {\dot z}^{\nu} \, .
\eeq
Let us note that the string contribution to the energy-momentum tensor,
\beq
T^{\mu \nu} = \frac{2}{\sqrt{g}}\, \frac{\delta S_m}{\delta g_{\mu \nu}} \,,
\eeq
reads
\beq
T^{\mu \nu} = \frac{\mu}{\sqrt{g}}\,\int d^2 \sigma\,e^{2\alpha\,\vphi}\, 
U^{\mu \nu}\,\delta^4(x -z(\sigma)) \,,
\eeq
where $\int d^4x \, \delta^4(x)=1$ and 
\beq
\label{car}
U^{\mu \nu} \equiv -\sqrt{\gamma} \, \gamma^{ab} \, \partial_a \, z^{\mu} \, 
\partial_b \, z^{\nu} = {\dot z}^{\mu} \, {\dot z}^{\nu} - z^{\prime\,\mu} \, 
z^{\prime \,\nu} 
\qquad \hbox{(in conformal gauge)}
\eeq
is the ``vertex operator'' for the interaction of the string with the 
gravitational field $g_{\mu \nu}$. [The vertex operators used here are the $x$-space 
counterparts of the $k$-space vertex operators used in quantum 
string theory, e.g. $\widehat{U}^{\mu \nu}(k) = 
\int d^2\sigma\,U^{\mu \nu}(z(\sigma))\,\exp(i\,k_\lambda\,z^\lambda(\sigma))$.]
The corresponding vertex operator for the 
interaction with the dilaton $\varphi$ is simply the trace $U \equiv g_{\mu 
\nu} \, U^{\mu \nu}$, while the one corresponding to the axion $B_{\mu \nu}$ 
is
\beq
V^{\mu \nu} \equiv - \epsilon^{ab} \, \partial_a \, z^{\mu} \, \partial_b 
\, z^{\nu} = {\dot z}^{\mu} \, z'^{\nu} - {\dot z}^{\nu} \, z'^{\mu} \, .
\eeq
The exact equation of motion of the string can be written (in any worldsheet 
gauge) as
\beq
\label{string}
0= \frac{\delta S_s}{\delta z^\mu} \equiv
\mu\,g_{\mu \nu}\,e^{2\alpha \vphi}\,
\pa_a (\sqrt{\gamma}\,\gamma^{a b}\,\pa_b z^\mu) + {\Phi}_\mu\,,
\eeq
where the quantity ${\Phi}_{\mu}$ is defined by
\beq
{\Phi}_{\mu} \equiv {\Phi}_{\mu}^{\varphi} + {\Phi}_{\mu}^h + {\Phi}_{\mu}^B\,,
\eeq
with
\bea
\label{f1}
{\Phi}_\mu^\vphi &=& 
\mu\,\alpha\,e^{2\alpha \vphi}\,U\,\pa_\mu \vphi-2\mu\,\alpha\,
e^{2\alpha \vphi}\,g_{\mu \alpha}\,U^{\alpha \beta}\,\pa_\beta \vphi\,,\\\vrbig
\label{f2}
{\Phi}_\mu^h &=& -\mu\,e^{2\alpha \vphi}\,g_{\mu \nu}\,\Gamma_{\alpha \beta}^\nu\,
U^{\alpha \beta} = 
\frac{\mu}{2}\,e^{2\alpha \vphi}\,U^{\alpha \beta}\,\pa_{\mu} h_{\alpha \beta} -
\mu\,e^{2\alpha \vphi}\,U^{\alpha \beta}\,\pa_\alpha h_{\beta \mu}\,,\\
\label{f3}
{\Phi}_\mu^B &=& \frac{\lambda}{2}\,V^{\alpha \beta}\,
H_{\mu \alpha \beta} =
\frac{\lambda}{2}\,V^{\alpha \beta}\,
\pa_\mu B_{\alpha \beta} + \lambda\,V^{\alpha \beta}\,
\pa_\alpha B_{\beta \mu}\,.
\eea
Let us emphasize that, while ${\Phi}_\mu^{\varphi}$ and ${\Phi}_\mu^B$ 
are well defined, spacetime and worldsheet covariant objects, 
${\Phi}_\mu^h$, by contrast, is not a covariantly defined 
object (but the full combination 
$\delta S_s/\delta z^\mu$ of Eq.~(\ref{string})  is a covariant object). 
It can be noted that the sum of the dilatonic and gravitational 
contributions (\ref{f1}), (\ref{f2}) 
simplify if they are expressed in terms of the string metric
$g_{\mu \nu}^s \equiv e^{2 \alpha \vphi}\,g_{\mu \nu}$ to which the string is directly 
coupled. Indeed, 
\beq
\label{phi+h}
{\Phi}_\mu^h + {\Phi}_\mu^{\vphi} = 
-\mu\,g^s_{\mu \nu}\,\Gamma_{\alpha \beta}^\nu[g^s_{\rho \sigma}]\,U^{\alpha \beta}=
\frac{\mu}{2}\,U^{\alpha \beta}\,\pa_{\mu} g^s_{\alpha \beta} -
\mu\,U^{\alpha \beta}\,\pa_\alpha g^s_{\beta \mu}\,.
\eeq
Except when otherwise specified, we shall henceforth work in the conformal gauge associated 
to the actual metric in which the string evolves (and not the 
conformal gauge associated to, say, a flat background metric $\eta_{\mu \nu}$). 
In this gauge the equations of motion of the string read 
\beq
\label{evo}
{\cal E}_\mu = 0, \quad \mbox{\small with } \quad 
{\cal E}_\mu \equiv \mu\,g^s_{\mu \nu}\,\eta^{a b} \,\pa_{a b} z^\nu + {\Phi}_\mu 
\equiv - \mu\,g^s_{\mu \nu}\,(\ddot{z}^\nu - z^{\nu\,\prime \prime}) + {\Phi}_\mu\,.
\eeq
When using this gauge, one must remember that the constraints (\ref{con}) 
[which read the same when written in terms of the string metric $g^s_{\mu \nu}$] 
involve the metric. These constraints read (in terms of the string metric) 
\beq
\label{constraints}
T_{a b}^s = 0, \quad \mbox{\small with} \quad  
T_{a b}^s \equiv g^s_{\mu \nu} (z) \, \partial_a \, 
z^{\mu} \, \partial_b \, z^{\nu} - \frac{1}{2} \, \eta_{ab} \, \eta^{cd} \, 
g^s_{\mu \nu} (z) \, \partial_c \, z^{\mu} \, \partial_d \, z^{\nu}\,.
\eeq
The constraints $T_{a b}^s = 0$ are preserved by the (gauge-fixed) evolution 
(\ref{evo}). Indeed, it is easy to check the identity 
\beq
\label{id}
\eta^{a b}\,\pa_a T_{b c}^s \equiv 
\pa_c z^\mu \,( g_{\mu \nu}^s\,\eta^{a b}\,\pa_{a b} z^\nu + {\Phi}_\mu^{\vphi}
+ {\Phi}_\mu^{h}) \equiv \pa_c z^\mu\,{\cal E}_\mu\,.  
\eeq
In the last step of Eq.~(\ref{id}) we used the algebraic identity 
$\pa_c z^\mu\,{\Phi}_\mu^B \equiv 0$. 
When the gauge-fixed equations of motion are satisfied, i.e. 
when $ {\cal E}_\mu =0$, the constraints satisfy
the conservation law $\eta^{a b}\,\pa_a T^s_{b c}=0$. 
This conservation law together with the algebraic identity 
$\eta^{a b}\,T^s_{a b} \equiv 0$ (i.e. $T^s_{00} = T^s_{11}$), 
ensures that if $T^s_{a b}$ vanishes on some initial slice $\tau = \tau_0$ , 
it will vanish everywhere on the worldsheet. This shows that the evolution 
equations (\ref{evo}) propagate only the physical, transverse 
degrees of freedom of the string.

Up to this point, we have made no weak-field approximation. In the following, we 
shall limit ourselves to working with formal perturbative expansions of the form 
(\ref{n2.1}), (\ref{n2.2}). When doing this, it is convenient to rewrite 
the string equations of motion (\ref{evo}) in the explicit form 
\beq
\label{n2.3}
{\cal E}_\mu = -\mu\,\eta_{\mu \nu}\,(\ddot{z}^\nu - z^{\prime \prime \nu}) 
+ {\cal F}_\mu\,,
\eeq
where the quantity ${\cal F}_\mu$ is defined as 
\beq
\label{n2.4}
{\cal F}_\mu \equiv {\Phi}_\mu + \Psi_\mu\,,
\eeq
with 
\beq
\label{n2.5}
 \Psi_\mu \equiv - \mu\,(g_{\mu \nu}\,e^{2\,\alpha\,\vphi} - \eta_{\mu \nu})\,
(\ddot{z}^\nu - z^{\nu \prime \prime})\,.
\eeq
In the linearized approximation, the complementary contribution $\Psi_\mu$ 
to the equations of motion read 
\beq
\Psi_\mu = -\mu\,(h_{\mu \nu} + 2\,\alpha\,\vphi\,\eta_{\mu \nu})\,
(\ddot{z}^\nu - z^{\nu \prime \prime}) + {\cal O}(G^2) \,.
\label{n2.6}
\eeq
The total contribution ${\cal F}_\mu$ to the explicit (non-covariant) 
string equations of motion is not 
a covariantly defined object, it is a non-covariant, pseudo-force density.
For the definition of a genuine, covariant 
force density see Ref.~\cite{BattyeCarter}, 
notably Eq.~(41) there. To simplify the language, 
we shall however call, in this paper, the non-covariant combination 
${\cal F}_\mu$ a ``force density'' (in the same way that when doing explicit 
calculations of the perturbative equations of motion of binary systems it is 
convenient to refer to the right-hand side of the equations 
of motion as a ``gravitational force''.)

Let us now write explicitly the weak-field approximation of the field equations 
deriving from the total action $S_f + S_s$. Let us recall that we assume 
the absence of external fields, so that we work with perturbative expansions of the 
form (\ref{n2.1}). 
When fixing the gauge freedom of the gravitational and axionic fields in the 
usual way ($g^{\alpha \beta} \, \Gamma_{\alpha \beta}^{\mu} = 0$; 
$\nabla^{\nu} 
\, B_{\mu \nu} = 0$), the field equations derived from $S_f$ read, at 
linearized order 
\bea
\label{e1}
\Box \vphi(x) &=& -4 \pi \, \int d^2 \sigma\, \Sigma^\vphi \, \delta^4(x -z(\sigma)) 
+ {\cal O}(G^2)\,,\\
\label{e2}
\Box h_{\mu \nu}(x) &=& -4 \pi \, \int d^2 \sigma\, \Sigma^h_{\mu \nu} \, \delta^4(x -z(\sigma)) + {\cal O}(G^2)\,,\\
\label{e3}
\Box B_{\mu \nu}(x) &=& -4 \pi \, \int d^2 \sigma\, \Sigma^B_{\mu \nu} \, 
\delta^4(x -z(\sigma)) + {\cal O}(G^2)\,,
\eea
where the corresponding linearized source terms are defined as
\beq
\label{source}
\Sigma^\vphi = \alpha G\,\mu\, U \,, \quad \quad  
\Sigma^h_{\mu \nu} = 4\,G\,\mu\,\widetilde{U}_{\mu \nu} \,, \quad \quad 
\Sigma^B_{\mu \nu} = 4\,G\, \lambda\,V_{\mu \nu} \,.
\eeq
Here, ${\widetilde U}_{\mu \nu} \equiv U_{\mu \nu} - \frac{1}{2} \, \eta_{\mu 
\nu} \, U$, and, in the present approximation, the vertex operators entering 
these source terms are simply $U_{\mu \nu} = {\dot z}_{\mu} \, {\dot z}_{\nu} 
- z'_{\mu} \, z'_{\nu}$, $U = \eta^{\mu \nu} \, U_{\mu \nu}$, $V_{\mu \nu} = 
{\dot z}_{\mu} \, z'_{\nu} - {\dot z}_{\nu} \, z'_{\mu}$, where we freely use 
the flat metric $\eta_{\mu \nu}$ to move indices. In the following, 
we shall consistently work to first order in $G$ only, and shall 
most of the time omit, to save writing, the indication of the 
${\cal O}(G^2)$ error terms.

The field equations (\ref{e1})--(\ref{e3}) are classically solved by introducing the four 
dimensional retarded Green function
\bea
\label{gret}
G_{\rm ret}(x-y) &=& \frac{1}{2 \pi}\,\theta(x^0-y^0)\,\delta((x-y)^2)\,; \\
\Box G_{\rm ret}(x-y) &=& -\delta^{(4)}(x-y)\,.
\eea
This ``retarded'' Green function incorporates the physical boundary condition 
of the non-existence of preexisting radiation converging from infinity toward 
the 
string source. The (unphysical) time-reverse of $G_{\rm ret}$ is the 
``advanced'' Green function
\bea
\label{gadv}
G_{\rm adv} (x-y) &= & \ \frac{1}{2\pi} \, \theta (-(x^0 - y^0)) \, \delta 
((x-y)^2)\,, \\
&= & \ G_{\rm ret} \, (y-x) \, . 
\eea
Let us consider, as a general model for Eqs. (\ref{e1})--(\ref{e3}), 
the generic field equation
\beq
\label{box}
\Box A(x) = -4 \pi \, \int d^2 \sigma\, \Sigma(\sigma) \, 
\delta^4(x -z(\sigma)) \,.
\eeq
Its most general classical solution reads
\beq
\label{class}
A_{\rm ret}(x) = +
4 \pi\,\int d\sigmap\,d\taup\,\Sigma(\sigmap,\taup)\,G_{\rm ret}(x-z(\sigmap,\taup))
+ A_{\rm ext}(x)
\,,
\eeq
where $A_{\rm ret}(x)$ is an ``external'' field, i.e. a generic homogeneous 
 solution of the field equations (generated by far away sources). 
As said above,  we assume in this work that $A_{\rm ext}(x)=0$. 

Applying the formula $\delta \, (F(\tau')) = {\displaystyle \sum_{\tau_0}} \,
\delta \, (\tau' - \tau_0) / \vert \partial \, F (\tau_0) / \partial \, \tau_0 
\vert$, where the sum runs over all the solutions $\tau_0$ of $F(\tau') = 0$, 
one can effectuate the integral over $\tau'$ in Eq.~(\ref{class}) 
with the result
\beq
\label{2.30}
A_{\rm ret}(x) = \int d\sigmap \left ( \frac{\Sigma(\sigmap,\taup)}{|\Omega \cdot \dot{z}|}
\right )_{{\big |}_{\taup = {\tau_{\rm ret}}}}\,.
\eeq
Here, we have defined $\Omega^{\mu} \, (x,\sigma',\tau') \equiv x^{\mu} - z^{\mu} \, 
(\sigma',\tau')$, and $\tau_{\rm ret} \, (x,\sigma')$ as being the retarded 
(i.e. such that $x^0 - z^0 \, (\tau_{\rm ret} (x,\sigma')) > 0$) 
solution in $\tau'$
of $\eta_{\mu \nu}\,\Omega^\mu(\tau')\,\Omega^\nu(\tau') = 0$. 
In the following, we use also the quantity $\partial_{\mu} \, 
A_{\rm ret}$ which, after using the formula
\beq
\partial_{\mu} \, \delta (F(x,\tau')) = \partial_{\mu} \, F \ \delta' 
(F(x,\tau')) = \frac{\partial_{\mu} \, F}{(\partial \, F / \partial \, \tau')} \, 
\frac{\partial \,\delta (F)}{\partial \, \tau'}
\eeq
and integrating by parts, can be written as
\beq
\label{der}
\pa_\mu A_{\rm ret}(x) = \int d\sigmap \left [\frac{1}{|\Omega \cdot \dot{z}|}\,
\frac{d}{d \taup}
\left ( \frac{\Omega_\mu\,\Sigma(\sigmap,\taup)}{\Omega \cdot \dot{z}}
\right )\right ]_{{\big |}_{\taup = {\tau_{\rm ret}}}}\,.
\eeq
The corresponding results for the advanced fields are
\beq
A_{\rm adv} (x) = \int d\sigmap \left 
( \frac{\Sigma(\sigmap,\taup)}{|\Omega \cdot \dot{z}|}
\right )_{{\big |}_{\taup = {\tau_{\rm adv}}}}\,,
\eeq
\beq
\partial_{\mu} \, A_{\rm adv} (x) =  
\int d\sigmap \left [\frac{1}{|\Omega \cdot \dot{z}|}\,\frac{d}{d \taup}
\left ( \frac{\Omega_\mu\,\Sigma(\sigmap,\taup)}{\Omega \cdot \dot{z}}
\right )\right ]_{{\big |}_{\taup = {\tau_{\rm adv}}}}\,,
\eeq
where $\tau_{\rm adv} (x^{\mu},\sigmap)$ is the advanced solution of 
$\eta_{\mu \nu}\,\Omega^\mu(\tau')\,\Omega^\nu(\tau') = 0$. 
Note that the scalar product $\Omega \cdot \dot z$ is {\it 
negative}\/ for $\tau' = \tau_{\rm ret}$ and {\it positive}\/ for $\tau' = 
\tau_{\rm adv}$.

\section{Perturbative on shell finiteness (and renormalizability) 
of the string self-interactions}
\label{sec3}
As said above, 
we consider the problem of a cosmic string interacting with its own, 
linearized, gravitational, 
dilatonic and axionic fields. The equations of motion of such a 
string read  
\beq
\mu\,\eta_{\mu \nu}\,(\ddot{z}^\nu - z^{\nu \prime \prime}) = 
{\cal F}_\mu(z) + {\cal O}(G^2)\,,\quad \quad 
\hbox{with} \,\,\,{\cal F}_\mu(z) = {\cal F}_\mu^{\rm lin}[
A^{\rm ret}(x),\pa A^{\rm ret}(x)]_{x = z}\,,
\label{n3.1}
\eeq  
where the explicit expression of the linearized ``force density'' 
${\cal F}_\mu^{\rm lin}$ is a linear functional of the 
(linearized) retarded fields $\vphi^{\rm ret}(x), h_{\mu \nu}^{\rm ret}(x), 
B_{\mu \nu}^{\rm ret}(x)$, 
\beq
{\cal F}_\mu^{\rm lin}[A^{\rm ret}, \pa A^{\rm ret}] 
= \Phi_\mu^{\vphi\, {\rm lin}}[\pa\vphi^{\rm ret}]
 + \Phi_\mu^{h\, {\rm lin}}[\pa h^{\rm ret}] + 
\Phi_\mu^{B\, {\rm lin}}[\pa B^{\rm ret}] + 
\Psi_\mu^{\rm lin}[\vphi^{\rm ret},h^{\rm ret}] \,,
\label{n3.2}
\eeq
where
\bea
{\Phi}_\mu^{\vphi \,\rm lin}[\pa \vphi^{\rm ret}] &=& 
\mu\,\alpha\,\eta_{\alpha \beta}\,U^{\alpha \beta}\,(\pa_\mu \vphi)
-2\mu\,\alpha\,\eta_{\mu \alpha}\,U^{\alpha \beta}\,(\pa_\beta \vphi)\,,
\nonumber \\\vrbig
{\Phi}_\mu^{h \,\rm lin}[\pa h^{\rm ret}] &=& 
\frac{\mu}{2}\,U^{\alpha \beta}\,(\pa_{\mu} h_{\alpha \beta}) -
\mu\,U^{\alpha \beta}\,(\pa_\alpha h_{\beta \mu})\,,\nonumber \\
{\Phi}_\mu^{B \,\rm lin }[\pa B^{\rm ret}] &=& 
\frac{\lambda}{2}\,V^{\alpha \beta}\,
(\pa_\mu B_{\alpha \beta}) + \lambda\,V^{\alpha \beta}\,
(\pa_\alpha B_{\beta \mu})\,, \nonumber \\\vrbig
\Psi_\mu^{\rm lin}[\vphi,h] &=& -\mu\,(h_{\mu \nu} + 2\,\alpha\,\vphi\,\eta_{\mu \nu}) 
\,(\ddot{z}^\nu - z^{\nu \prime \prime})\,.
\label{n3.3}
\eea
The right-hand side of Eqs.~(\ref{n3.3}) is obtained by inserting 
the retarded fields $A_{\rm ret}(x)$, Eq.~(\ref{2.30}), and their first derivatives, 
$\pa_\mu A_{\rm ret}(x)$, Eq.~(\ref{der}), and by evaluating 
the result at a point $x^\mu = z^\mu$ on the string worldsheet. The sources 
of the fields are given in terms of the string dynamics by 
Eqs.~(\ref{source}). Note that ${\cal F}_\mu$ is a non-local functional of 
the string worldsheet whose support is the intersection of the worldsheet 
with the past light cone with vertex at the point $z^\mu$.

As in the case of a self-interacting point particle, the force 
${\cal F}_{\mu} (x=z)$ is infinite because of the divergent contribution 
generated when the source point $z^{\mu} (\tau' , \sigma')$ coincides with the 
field point $x^{\mu} = z^{\mu} (\tau , \sigma)$. It was emphasized long ago by 
Dirac \cite{Dirac38}, in the case of an electron moving in its own 
electromagnetic field, that this problem can be cured by renormalizing the 
mass, 
thereby absorbing the divergent part of the self-force. More precisely, Dirac 
introduced a cut-off radius $\delta$ around the electron and found a 
corresponding (ultraviolet divergent) 
self-force ${\cal F}^{\mu} (\delta) = -(e^2 / 2 \, \delta) \, 
{\ddot z}^{\mu} + {\cal F}_{\rm R}^{\mu}$ where ${\cal F}_{\rm R}^{\mu}$ is a finite 
(renormalized) contribution. If the mass of the electron plus its 
$\delta$-surrounding depends on $\delta$ according to
\beq
\label{dirac1}
m(\delta) = m_{\rm R} - \frac{e^2}{2 \, \delta} \, ,
\eeq
where $m_{\rm R}$ denotes a finite, ``renormalized'' mass, the ultraviolet divergent 
equations of motion $m(\delta) \, {\ddot z}^{\mu} = {\cal F}^{\mu} (\delta)$ 
give the finite result $m_{\rm R} \, {\ddot z}^{\mu} = {\cal F}_{\rm R}^{\mu}$. Note that 
the $\delta$-dependence of $m(\delta)$ (for a fixed $m_{\rm R}$) is compatible with 
the idea that $m(\delta)$ represents the total mass-energy of the particle 
plus 
that of the electromagnetic field contained within the radius $\delta : 
m(\delta_2) - m(\delta_1) = + \int_{\delta_1}^{\delta_2} d^3 x \, (8\pi)^{-1} 
\, 
(e/r^2)^2$. Dirac also found that the remaining finite force was given by 
(using a proper-time normalization of $\tau : {\dot z}^2 = \eta_{\mu \nu} \, 
dz^{\mu} / d\tau \, dz^{\nu} / d\tau = -1$) 
the sum of the external force ${\cal F}_{\rm ext}^\mu$ and of a finite 
``reactive'' self-force ${\cal F}_{\rm reac}^\mu$,
\beq
\label{dirac2}
{\cal F}_{\rm R}^\mu = {\cal F}_{\rm ext}^\mu + {\cal F}_{\rm reac}^\mu\,,
\quad {\cal F}_{\rm ext}^{\mu} = {F}_{\rm ext}^{\mu \nu}\,\dot{z}_\nu\,, \quad  
{\cal F}_{\rm reac}^{\mu} \equiv \frac{1}{2} \, (F_{\rm 
ret}^{\mu \nu} - F_{\rm adv}^{\mu \nu}) \, {\dot z}_{\nu} = \frac{2}{3} \, e^2 
({\tdot z}^{\mu} + (\dot z \cdot \tdot z) \, {\dot z}^{\mu}) \, .
\eeq
The analogous problem for self-interacting cosmic strings has been studied 
by Lund and Regge \cite{LundRegge} and 
Dabholkar and Quashnock \cite{DQ90} for the coupling to the axion field (see 
also \cite{BS96}), by Copeland, Haws and Hindmarsh \cite{CHH90} for the 
couplings to gravitational, dilatonic and axionic fields 
(and by Carter \cite{C97} for the couplings 
to electromagnetic fields). There is, however, a subtlety in the calculation 
of the renormalization of the string equations of motion which led 
Ref.~\cite{CHH90} (and us, in the first version of this work) to misinterpret 
their results, and propose incorrect values of the renormalizations 
of the string tension due to gravitational and dilatonic self-interactions. 
Our realization of this subtlety, was triggered by the work of Carter 
and Battye \cite{CB}, who were the first to get the correct renormalization 
of $\mu$ under self-gravitational effects, in $4$-dimensions, by using a covariant 
approach to string dynamics \cite{C89+93}, \cite{BattyeCarter}. We then 
obtained \cite{BDletter}  the correct renormalizations of $\mu$ under all three 
fields, and in an arbitrary spacetime dimension~\footnote{
Only the leading divergence was treated when $n > 4$.}, by an effective action 
approach. The subtly which makes it delicate (but not impossible) 
to derive the correct renormalization of $\mu$ when working 
(as Ref.~\cite{CHH90} and the present paper) directly with the equations of motion, 
at first order in $G$, and without adding external fields, is the following. In such a context, the {\it perturbative} string equations of motion (\ref{n2.2}) 
imply that $\ddot{z}^\mu - z^{\mu \prime \prime}$ is of order $G$, so that any 
first-order renormalization of the tension, $\mu = \mu^0 + G\,\mu^1 + 
{\cal O}(G^2)$, corresponds only to {\it second order} 
contributions ($G\,\mu^1\,(\ddot{z}_\mu - z_{\mu}^{\prime \prime}) 
\sim G^2\,\mu^1\,{\cal F}_\mu^1 = {\cal O}(G^2)$) which are formally negligible 
at order ${\cal O}(G)$ and, therefore, cannot be unambiguously read off such a 
first-order calculation. In other words, a first-order treatment without external 
fields can only prove that the string equations 
of motion are {\it renormalizable} by checking their {\it on
(perturbative) shell finiteness} (i.e. the fact that all formally 
divergent first-order contributions vanish when using the zeroth-order string 
equations of motion  $\ddot{z}^\mu - z^{\mu \prime \prime} = 0 
+ {\cal O}(G)$), but cannot, by themselves, unambiguously determine  
the renormalization of the string tension. For instance, the finding 
of Quashnock and Spergel \cite{QuashnockSpergel90} 
that the self-gravity effects vanish upon using the 
zeroth-order equations of motion to evaluate the first-order terms in Eq.~(\ref{n2.2}) 
prove that they are renormalizable, but does not allow one to conclude 
that the self-gravity contribution to the tension renormalization, 
$\delta_g \mu$ vanishes. [It happens that $\delta_g \mu$ vanishes 
in $4$-dimensions \cite{CB}, but this vanishing is an ``accident'' 
which does not hold in other spacetime dimensions  \cite{BDletter}.] 
To be able to determine the value of the renormalization of $\mu$ 
one must go beyond a zero-background, first-order ``on-shell'' 
treatment of the string equations of motion. Essentially, one must 
work with a form of the string equations of motion which allow for the unambiguous 
introduction of an ``external force'' acting on the string. This is the case 
of the covariant-force formalism of Ref.~\cite{CB}, as well as of the 
effective-action formalism of Ref.~\cite{BDletter} (where an extra force 
would mean an additional contribution to the total action.) 
In the present work, we do not really need the explicit value 
of the tension renormalization. We only need to check the 
{\it renormalizability} of the perturbative string equations of motion, 
i.e. the fact that all infinities vanish ``on (zeroth-order) shell''. 
To end up with clearer results, we shall, however, present a treatment 
in which the correct renormalization appear (because this treatment 
is action-based), and we shall renormalize them away by using, as external input, 
the results of Refs.~\cite{CB}, \cite{BDletter}.

Our starting point will be the explicit form, Eqs.~(\ref{n3.1})--(\ref{n3.3}), 
of the conformal-gauge, variational equations of motion 
$\delta S_s/\delta z^\mu = 0$. [We shall check below that the conformal-
gauge constraints (\ref{con}) (written with the full divergent 
metric) do not contain any divergent contributions (at linear order in $G$).]
To give a meaning to Eqs.~(\ref{n3.3}) when $x \rightarrow z(\tau, \sigma)$ 
we formally 
introduce an ultraviolet cutoff $\delta_c$ in 
the $\sigma'$-integration giving the retarded fields and their derivatives, 
i.e. we replace the integral over a full period of 
$\sigma'$, $\int_{\sigma_0}^{\sigma_0 + L} d\sigma'$, on the right-hand side of 
Eqs.~(\ref{2.30}) and (\ref{der}) 
by $\int_{\sigma_0}^{\sigma - \delta_c} d\sigma' + \int_{\sigma + 
\delta_c}^{\sigma_0 + L} d\sigma'$. 
Later in this paper, we shall use a different way to introduce an ultraviolet cutoff, 
namely dimensional regularization. Dimensional regularization has the advantage of 
always keeping Lorentz invariance manifest. We have checked that both methods give the same 
results (see Appendix). In this section, we use the less sophisticated 
$\delta_c$-cutoff approach which allows a more direct comparison with other 
results in the literature. 

We then need the expansions in powers of 
$\sigma' - \sigma$ and $\tau' - \tau$ of all the quantities entering 
Eqs.~(\ref{2.30}), (\ref{der}):
\bea
\Omega_{\mu}(\taup,\sigmap) &\simeq& 
-\,(\sigmap - \sigma)\,z_\mu^\prime - (\taup-\tau)\,\dot{z}_\mu 
-\frac{1}{2}(\sigmap-\sigma)^2\,z_\mu^{\prime \prime} 
-\frac{1}{2}(\taup-\tau)^2\,\ddot{z}_\mu  \nonumber \\
&& -\,(\sigmap-\sigma)\,(\taup-\tau)\,\dot{z}^\prime_\mu \,, \vrsmall \\
(\Omega \cdot z)(\taup,\sigmap)
&\simeq& -\,(\taup-\tau)\,\dot{z}^2 - \frac{3}{2}(\taup-\tau)^2\,(\dot{z}\cdot \ddot{z})
+\frac{1}{2}(\sigmap-\sigma)^2\,(z^{\prime \prime}\cdot \dot{z}) \nonumber \\
&& + \,(\sigmap-\sigma)\,(\taup-\tau)\,(\ddot{z}\cdot z^\prime)\,, \vrsmall \\
\dot{z}_\mu(\taup,\sigmap) &\simeq& \dot{z}_\mu + (\taup-\tau)\,\ddot{z}_\mu + 
(\sigmap-\sigma)\,\dot{z}_\mu^\prime \,.
\eea
At the order needed to extract the divergent part of the integrals (\ref{2.30}), 
(\ref{der}) (we shall use a more efficient tool below to extract 
the more complicated 
finite reactive part) it is enough to use 
\beq
\label{n3.5}
\tau_{\rm ret}(z^\mu, \sigma^\prime) = \tau - |\sigma - \sigma^\prime| + 
{\cal O}(|\sigma - \sigma^\prime|^2)\,,
\eeq
for the retarded solution of $\eta_{\mu \nu}\,\Omega^\mu(\tau^\prime)\, 
\Omega^\nu(\tau^\prime) = 0$. Inserting these results in Eqs.~(\ref{2.30}), 
(\ref{der}) we get 
\beq
\label{n3.6}
A_{\rm ret}(z) = \frac{1}{|\dot{z}^2|}\, \log \left (\frac{1}{\delta_c}
\right )\,[2\,\Sigma ] \quad + \quad \hbox{finite terms}\,,
\eeq
\bea
\label{delder}
\pa_\mu A_{\rm ret}(z) &&= 
\frac{1}{(\dot{z}^2)^2}\,\log \left (\frac{1}{\delta_c} \right )\,
\left [ - \Sigma\,\ddot{z}_\mu + \Sigma\,z_{\mu}^{\prime \prime} + 4\Sigma\,z_\mu^\prime\,
\left ( \frac{z^{\prime \prime}\cdot z^\prime}{\dot{z}^2}\right)
+4\Sigma\,\dot{z}_\mu\,\left (\frac{\ddot{z}\cdot \dot{z}}{\dot{z}^2}\right )  \right .
\nonumber \\
&& \left . + 2 \Sigma^\prime\,z^\prime_\mu - 2\dot{\Sigma}\,\dot{z}_\mu \right ]
\quad + \quad \hbox{finite terms}\,.
\eea
The rather complicated-looking terms proportional to $(z' \cdot z'') / {\dot 
z}^2$ and $(\dot z \cdot \ddot z) / {\dot z}^2$ in Eq.~(\ref{delder}) are, actually, 
``connection'' terms linked to the fact that the source $\Sigma$ is a 
worldsheet density (conformal weight 2) rather than a worldsheet scalar 
(conformal weight 0). Let us associate to each source $\Sigma$ a corresponding 
worldsheet scalar $S$, also denoted $\widehat{\Sigma}$, defined by
\beq
S \equiv \widehat{\Sigma}\equiv \frac{1}{\sqrt{\gamma}} \ \Sigma \, .
\eeq
Here $\sqrt{\gamma} = (-\det \, \gamma_{ab})^{1/2}$ is the area-density $dA / 
d^2 \sigma$, which reads, in conformal gauge: $\sqrt{\gamma} = z'^2 = - 
{\dot z}^2$. One needs also to introduce the invariant ultraviolet cutoff 
$\delta \equiv \gamma^{1/4}\,\delta_c \equiv (z^{\prime 2})^{1/2}\,\delta_c$
associated to the ``coordinate cutoff'' $\delta_c$. 
(In Sec.~\ref{sec4} below and in the Appendix, we shall use a dimensional 
regularization method where the cutoff parameter $\epsilon = 4-n$, 
and the renormalization scale $\Delta_{\rm R}$, are automatically Lorentz invariant). 
Then Eqs.~(\ref{n3.6}), (\ref{delder}) simplify to
\beq
\label{n3.8}
A_{\rm ret} (z) = 
\log \left (\frac{1}{\delta}\right ) \, \left [2 \, S \right ] \quad + \quad 
\hbox{finite terms}\,,
\eeq
\beq
\label{3.8}
\partial_{\mu} \, A_{\rm ret} (z) = \frac{1}{\sqrt{\gamma}} \, 
\log \left (\frac{1}{\delta}\right) \, 
\left [-S \, {\ddot z}_{\mu} + S \, z''_{\mu} - 2 \, \dot S \, 
{\dot z}_{\mu} + 2 \, S' z'_{\mu}\right ] \quad + \quad \hbox{finite terms} .
\eeq
The result (\ref{3.8}) for the regularized field derivative agrees with the results
of Ref.~\cite{CHH90}, as well as  with the geometric 
prescription given in \cite{C97}. 
As a check on the above results one can verify that the divergent parts 
satisfy
\bea
&& \frac{\partial}{\partial \tau} \, A_{\rm ret} (z) = {\dot z}^{\mu} \, 
\partial_{\mu} \, A_{\rm ret} (z) \,,\\
&& \frac{\partial}{\partial \sigma} \, A_{\rm ret} (z) = z^{\mu\,\prime} \, 
\partial_{\mu} \, A_{\rm ret} (z) \, .
\eea
To check these links one must use the following consequence of the conformal 
gauge constraints $0 = T_{ab}^E \equiv g_{\mu \nu} (z) \, \partial_a \, 
z^{\mu} \, \partial_b \, z^{\nu} - \frac{1}{2} \, \eta_{ab} \, \eta^{cd} \, 
g_{\mu \nu} (z) \, \partial_c \, z^{\mu} \, \partial_d \, z^{\nu}$ 
(written here in terms of the Einstein metric):
\beq
\label{conf}
0 = \eta^{bc}\,\pa_b\, T_{ca}^E \equiv g_{\mu \nu} (z) \, \partial_a \, 
z^{\mu} \, \eta^{cd} \, (\partial_{cd} \, z^{\nu} + \Gamma_{\alpha 
\beta}^{\nu} \, \partial_c \, z^{\alpha} \, \partial_d \, z^{\beta}) \, .
\eeq
As we assume everywhere in the paper a flat gravitational background 
$\eta_{\mu \nu}$, Eq.~(\ref{conf}) implies 
\beq
\label{3new}
{\dot z}^{\mu} ({\ddot z}_{\mu} - 
z''_{\mu}) = {\cal O}(h_{\alpha \beta})\,, \quad 
\quad z'^{\mu} ({\ddot z}_{\mu} - z''_{\mu})= {\cal O}(h_{\alpha \beta})\,,
\eeq
so that tangential projections of ${\ddot z}^{\mu} - z^{\mu\,\prime \prime}$
can be consistently neglected in first-order contributions such as 
Eqs.~(\ref{f1})--(\ref{f3}) even if one is working ``off-shell''.

Because of the logarithmic divergence entering Eq.~(\ref{3.8}) we need to introduce, 
besides the invariant ultraviolet cutoff scale $\delta$ (which can be thought of as the 
width of the cosmic string), an arbitrary, finite, renormalization length 
scale $\Delta_{\rm R}$. Then, we can {\it define} precisely 
the ``infinite parts'' (IP) of $A_{\rm ret}(z)$ and 
$\partial_{\mu} \, A_{\rm ret} (z)$, i.e. the parts which blow up when $\delta 
\rightarrow 0$, by replacing in Eqs.~(\ref{n3.8}), (\ref{3.8}) 
the logarithm by $\log \, (\Delta_{\rm R} 
/ \delta)$, and by discarding any other finite contribution. To apply this 
definition to the three fields $\varphi$, $h_{\mu \nu}$, and $B_{\mu \nu}$, we 
need to use the corresponding sources, Eq.~(\ref{source}). For instance, we have 
\bea
\mbox{IP}(h_{\mu \nu}^{\rm ret}(z))  &=& \frac{1}{\sqrt{\gamma}}\,2 \Sigma_{\mu \nu}^h\,\log\left 
(\frac{\Delta_R}{\delta} \right ) = 
8 G\,\mu\,\frac{1}{\sqrt{\gamma}}\, \tilde{U}_{\mu \nu} \, \log \left 
(\frac{\Delta_R}{\delta} \right )\,, \\
\mbox{IP}(\vphi^{\rm ret}(z)) &=& -4\alpha\,G\,\mu\,\log \left 
(\frac{\Delta_R}{\delta} \right )\,.
\eea
Using the easily verified 
identities satisfied by the vertex operators,
\bea
\label{3.14}
&& U_{\mu \nu} \, \widetilde{U}^{\mu \nu} \equiv U_{\mu \nu} \, U^{\mu \nu} - 
\frac{1}{2} \, U^2 = 0\,, \quad U_{\mu \sigma} \, \widetilde{U}^{\nu \sigma} = 0 \,,
\quad U_{\mu \nu} \, {\dot{\widetilde U}}^{\mu \nu} = 0 \, , \\
&&V_{\mu \nu} \, V^{\mu \nu} = -2 ({\dot z}^2)^2\,, \quad 
\dot{V}_{\mu \nu} \, \dot{V}^{\mu \nu} = -2\ddot{z}^2\,\dot{z}^2 
+ 2\dot{z}^2\,\dot{z}^{\prime\,2} -4(\ddot{z} \cdot \dot{z})^2
-4(\dot{z}^\prime \cdot \dot{z})^2\,,
\label{3.15}
\eea
we first see easily that the divergent contributions, 
$\mbox{IP}(h_{\mu \nu})(\dot{z}^\mu\,\dot{z}^\nu + 
{z}^{\mu \prime}\,{z}^{\nu \prime})$ and $\mbox{IP}(h_{\mu \nu})\,\dot{z}^\mu\,
{z}^{\nu \prime}$,
to the constraints (\ref{con}) vanish. The use of the identities 
(\ref{3.14}), (\ref{3.15}) allows also to simplify the expression of the terms 
linear in the field derivatives entering Eqs.~(\ref{n3.3}). We obtain 
\bea
&& \mbox{IP}(\pa_\mu \vphi) = 
-2\alpha\,G\,\mu\,(\ddot{z}_\mu - z^{\prime \prime}_\mu )\,\frac{1}{\dot{z}^2}\,
\log \left (\frac{\Delta_{\rm R}}{\delta} \right)\,, \\
&& \mbox{IP}(U^{\alpha \beta}\, \pa_\alpha h_{\beta \mu}) =
-8G\,\mu\,(\ddot{z}_\mu - z^{\prime \prime}_\mu )\,
\log \left (\frac{\Delta_{\rm R}}{\delta} \right)\,,\\
&& \mbox{IP}(U^{\alpha \beta}\, \pa_\mu h_{\alpha \beta }) = 0\,,\\
&& \mbox{IP}(V^{\alpha \beta}\,\pa_\mu B_{\alpha \beta}) = 8G\,\lambda\,
(\ddot{z}_\mu - z^{\prime \prime}_\mu )\,
\log \left (\frac{\Delta_{\rm R}}{\delta} \right)\,, \\
&& \mbox{IP}(V^{\alpha \beta}\,\pa_\alpha B_{\beta \mu}) = -8G\,\lambda\,
(\ddot{z}_\mu - z^{\prime \prime}_\mu )\,
\log \left (\frac{\Delta_{\rm R}}{\delta} \right)\,. 
\eea
We have now in hands all the results needed to derive the infinite 
contributions to the right-hand side 
of the string equations of motion (\ref{n3.1}). More precisely, 
one obtains for each separate contribution in  Eq.~(\ref{n3.2}) 
\bea
\label{3.20}
\mbox{IP}({\Phi}_\mu^\vphi) &=& -4\alpha^2\,G\,\mu^2\,(\ddot{z}_\mu - 
z_{\mu}^{\prime \prime})\,
\log \left (\frac{\Delta_{\rm R}}{\delta} \right)\,,\\   
\label{3.21}
\mbox{IP}({\Phi}_\mu^h) &=& 8G\,\mu^2\,(\ddot{z}_\mu - z_{\mu}^{ \prime \prime})\,
\log \left (\frac{\Delta_{\rm R}}{\delta} \right) \,,\\
\label{3.22}
\mbox{IP}({\Phi}_\mu^B) &=& -4G\,\lambda^2\,(\ddot{z}_\mu - z_{\mu}^{\prime \prime})\,
\log \left (\frac{\Delta_{\rm R}}{\delta} \right) \,, \\
\mbox{IP}(\Psi^\vphi_\mu) &=& 
 8 \alpha^2\,G\,\mu^2 ( 
\ddot{z}_\mu -z_\mu^{\prime \prime})\,
\log \left (\frac{\Delta_{\rm R}}{\delta} \right )\,,\\
\mbox{IP}(\Psi^h_\mu) &=&  -8 G\,\mu^2 ( 
\ddot{z}_\mu -z_\mu^{\prime \prime})\,
\log \left (\frac{\Delta_{\rm R}}{\delta} \right )\,.
\eea
Adding up all the terms leads to 
\beq
\mbox{IP}({\cal F}_\mu) = C\,(\ddot{z}_\mu - z_{\mu}^{\prime \prime})\,
\log \left (\frac{\Delta_{\rm R}}{\delta} \right )\,,
\label{n3.10}
\eeq
with 
\beq
C = + 4\alpha^2\,G\,\mu^2 - 4\,G\,\lambda^2\,.
\label{n3.11}
\eeq
The crucial point in the result (\ref{n3.10}) is that the 
divergent contribution to the equations of motion is proportional 
to the zeroth-order equations of motion. In our present 
perturbative treatment the formally infinite contribution 
(\ref{n3.10}) is of second order in $G$ and can be ignored. As we 
said above, this property of perturbative on shell finiteness 
of the equations of motion proves their renormalizability 
but cannot, by itself, determine the physically correct value of the 
renormalization of $\mu$. At this point, we can, however, use 
the results of Ref.~\cite{BDletter}, where we showed that the ``bare'' 
(regularized but not renormalized) string tension $\mu(\delta)$ 
appearing in the original ultraviolet-divergent 
action must depend on the UV cutoff $\delta$ according to 
\beq
\mu(\delta) = \mu_R + C\,\log \left (\frac{\Delta_{\rm R}}{\delta} \right )\,,
\label{n3.28}
\eeq
where $\mu_R$ is the finite, renormalized tension, and where 
the (``beta function'') coefficient $C$ is precisely given by 
Eq.~(\ref{n3.11}). [$C$ contains only contributions coming 
from dilatonic, $C_\vphi = +4\,\alpha^2\,G\,\mu^2$, and axionic, 
$C_B = - 4\,G\,\lambda^2$, self-interactions. The gravitational 
contribution vanishes (in $4$-dimensions) \cite{CB}, \cite{BDletter}.]

Let us define the ``renormalized'' value of any ``bare'' (i.e. cutoff-dependent), 
logarithmically divergent, quantity $Q(\delta)$ as its ``finite part'' (FP), 
i.e. the difference between $Q(\delta)$ and its ``infinite part'' 
(defined above as the term $\propto \log(\Delta_R/\delta)$ in $Q(\delta)$ )
\beq
\label{3.26}
{Q}^{\rm R} \equiv \mbox{FP} \, (Q(\delta)) \equiv \lim_{\delta \rightarrow 
0} \, \left [Q(\delta) - \mbox{IP} \, (Q (\delta))\right ] \, .
\eeq 
Using this definition, and formally inserting Eq.~(\ref{n3.28}) into the 
bare equations of motion (\ref{n3.1}), namely 
\bea
\label{nn1}
+\mu(\delta)\,\eta_{\mu \nu}\, 
(\ddot{z}^\nu - z^{\nu\,\prime \prime}) &=& \left [\mu_R + C\, 
\log \left (\frac{\Delta_{\rm R}}{\delta} \right )\right ]\,
\eta_{\mu \nu}\,(\ddot{z}^\nu - z^{\nu \prime \prime}) \nonumber \\
&=&{\cal F}_\mu(\delta) = 
\mbox{IP}({\cal F}_\mu(\delta)) + {\cal F}_\mu^R  \nonumber \\
\label{nn2}
&=& C\,(\ddot{z}_\mu - z_{\mu}^{\prime \prime})\,
\log \left (\frac{\Delta_{\rm R}}{\delta} \right ) + {\cal F}_\mu^R
+ {\cal O}(G^2)\,,
\label{nn3}
\eea
we see that the terms proportional to $\log \Delta_R/\delta$ 
coming from the renormalization of $\mu(\delta)$ and 
those coming from the renormalization of ${\cal F}_\mu(\delta)$
are identical (even if we were working 
off shell), so that the equations of motion can be 
rewritten in the renormalized form
\beq
\label{ren}
\mu_R\,\eta_{\mu \nu}\,(\ddot{z}^\nu - z^{\nu \prime \prime}) = 
{\cal F}_\mu^R+ {\cal O}(G^2)\,.
\eeq
This simplification between the same $C\,\log(\Delta_R/\delta)\,
(\ddot{z}_\mu - z_{\mu}^{ \prime \prime})$ contributions on both sides of the equations of motion 
is due to the fact that we have been working with the direct, Euler-Lagrange 
variational equations $\delta S_s/\delta z^\mu$, i.e. with a form of the equations 
of motion which is ready to receive an additional ``external force'' 
$\delta S^\prime/\delta z^\mu$, as the variational derivative of an additional piece 
$S^\prime$ in the action. Had we worked with another form of the equations of motion, say 
\beq
\widetilde{\cal E}^\mu \equiv g_s^{\mu \nu}\,
\frac{\delta S_s}{\delta z^\nu} \equiv g^{\mu \nu}\,
e^{-2 \alpha \vphi}\,{\cal E}_\nu \equiv - \mu(\delta)\,
(\ddot{z}^\mu - z^{\mu \prime \prime}) + \widetilde{\Phi}^\mu \,,
\eeq
with $ \widetilde{\Phi}^\mu \equiv g^{\mu \nu}\,e^{-2 \alpha \vphi}\,\Phi_\nu$, 
the infinite part of the {\it linearized} field-contribution 
$\widetilde{\Phi}^{\mu \,\rm lin}[\pa_\mu A_{\rm ret}]$ would have been 
identical to $\eta^{\mu \nu}\,{\Phi}_{\nu}^{\rm lin}[\pa_\mu A_{\rm ret}]$, 
with ${\Phi}_{\nu}^{\rm lin}$ given in Eq.~(\ref{n3.3}) above.
In such a case, Eqs.~(\ref{3.20})--(\ref{3.22}), show that the infinite 
part of $ \widetilde{\Phi}^{\mu\,\rm lin}$ would not have matched the infinite 
contribution $\mbox{IP}(\mu(\delta))\,(\ddot{z}^\mu - z^{\mu \prime \prime}) 
= C\,\log (\Delta_R/\delta)\,(\ddot{z}^\mu - z^{\mu \prime \prime})$.
This apparent discrepancy is, however, not at all a sign of inconsistency 
of the type of non-covariant perturbative equations of motion 
we have been using. Either one works on shell, and all the formally 
infinite terms can be consistently neglected as being of order $G^2$, 
or one introduces an additional mechanical interaction of the string, 
e.g. through the addition of a new piece 
$S^\prime(z^\mu, \dots)$ in the action, in which case the zeroth-order string 
``mass shell'' is modified, and we must take into account the new 
infinite terms coming from the extra contribution 
$g^{\mu \nu}\,e^{-2 \alpha \vphi}\,\delta S^\prime/ \delta z^\nu$, 
in which $g^{\mu \nu}\,e^{-2 \alpha \vphi}\, = \eta^{\mu \nu} - 
\mbox{IP}(h^{\mu \nu}) - 2\alpha\,\eta^{\mu \nu}\,
\mbox{IP}(\vphi) + \hbox{finite}$.

Let us finally note that the logarithmic renormalizations (\ref{n3.28}), 
(\ref{3.26})
introduce a dependence of the renormalized quantities 
upon an arbitrary, renormalization length scale $\Delta_R$. [By definition, 
the bare (regularized) quantities $\mu(\delta)$, $Q(\delta)$, 
do not depend on the choice of $\Delta_R$.] For instance, we see from 
Eqs.~(\ref{n3.10}), (\ref{n3.28}) that 
\bea
\label{3.31}
&& \mu_{\rm R} (\Delta'_{\rm R}) = \mu_{\rm R} (\Delta_{\rm R}) - C \, \log 
\left (\frac{\Delta'_{\rm R}}{\Delta_{\rm R}}
\right ) \,, \\
&& {\cal F}_\mu^R(\Delta'_{\rm R}) = {\cal F}_\mu^R(\Delta_{\rm R}) -
C\,(\ddot{z}_\mu - z_\mu^{\prime \prime})\,\log 
\left (\frac{\Delta'_{\rm R}}{\Delta_{\rm R}}\right )\,.
\label{n3.31}
\eea
It is however, easily seen that the content of the renormalized equations 
of motion (\ref{ren}) is left invariant (at first order in the field 
couplings) under a change of $\Delta_{\rm R}$. [This invariance still 
holds in presence of an additional (finite) contribution 
$\delta S^\prime/ \delta z^\mu$ to the equations of motion.]
As we work only to first order in the field couplings, note 
that the quantity $\mu$ appearing in $C$, Eq.~(\ref{n3.28}), can formally 
be considered as being 
a renormalized value, rather than the bare one, thereby leading to the 
renormalization group equation $\partial \, \mu_{\rm R} / \partial \, \log \, 
\Delta_{\rm R} = -C (\mu_{\rm R})$. [The non-renormalizability of the gravitational 
interaction makes it delicate to extend this argument to higher orders in $G$. 
By contrast, if we consider only a canonically normalized axionic field, with 
coupling $\sqrt{G}\,\lambda = \sqrt{\pi/2}\,f_a$, $C$ does not depend on $\mu$ and the 
first-order renormalization result is exact.]

Finally, we note that the axionic contribution  $C_{B} =-4G\,\lambda^2$
to $C$ agrees with the result of previous dynamical calculations 
\cite{LundRegge},\cite{CHH90}, \cite{QuashnockSpergel90}, \cite{BS95} and \cite{BS96}, 
while the dilatonic contribution $C_{\varphi} = +4\alpha^2\,G\,\mu^2$ 
disagrees with Ref.~\cite{CHH90} which proposed a vanishing dilatonic contribution 
$C_\vphi$.
 
\section{Renormalized force density and the local back-reaction approximation}
\label{sec4}
\subsection{Renormalized equations of motion}
\label{subsec4.1}
In the previous section we have shown that the perturbative equations of motion 
(in absence of external fields) could be written, at first order in $G$, 
in the renormalized form 
\beq
\mu_R\,\eta_{\mu \nu}\,(\ddot{z}^\nu - z^{\nu \prime \prime}) = 
{\cal F}_\mu^{\rm R lin}[\pa A_{\rm ret}] + 
{\cal O}(G^2)\,,
\label{N4.1}
\eeq
where the R.H.S. is the sum of three renormalized contributions 
\beq
{\cal F}_\mu^{\rm R\, lin}[\pa A_{\rm ret}] = 
{\cal F}_\mu^{\rm \vphi\, R\, lin}[\pa \vphi_{\rm ret}] + 
{\cal F}_\mu^{\rm h\, R\, lin}[\pa  h_{\rm ret}]
+ {\cal F}_\mu^{\rm B \,R\, lin}[\pa B_{\rm ret}]\,,
\label{n4.1}
\eeq
with 
\bea
\label{rforce1}
{\cal F}_\mu^{\vphi \rm R\, lin}(\pa \vphi) &=& 
\mu\,\alpha\,\eta_{\alpha \beta}\,U^{\alpha \beta}\,(\pa_\mu \vphi)^{\rm R}
-2\mu\,\alpha\,\eta_{\mu \alpha}\,U^{\alpha \beta}\,(\pa_\beta \vphi)^{\rm R}\,,\\\vrbig
\label{rforce2}
{\cal F}_\mu^{h \rm R\, lin}(\pa h) &=& 
\frac{\mu}{2}\,U^{\alpha \beta}\,(\pa_{\mu} h_{\alpha \beta})^{\rm R} -
\mu\,U^{\alpha \beta}\,(\pa_\alpha h_{\beta \mu})^{\rm R}\,,\\
\label{rforce3}
{\cal F}_\mu^{B \rm R\, lin }(\pa B) &=& 
\frac{\lambda}{2}\,V^{\alpha \beta}\,
(\pa_\mu B_{\alpha \beta})^{\rm R} + \lambda\,V^{\alpha \beta}\,
(\pa_\alpha B_{\beta \mu})^{\rm R}\,.
\eea
Here $(\pa_\mu A_{\rm ret})^R$, with $A_{\rm ret} = (\vphi^{\rm ret}, 
h_{\mu \nu}^{\rm ret}, B_{\mu \nu}^{\rm ret})$, denotes, as defined 
by Eq.~(\ref{2.30}), the finite part of the logarithmically 
divergent retarded integral (\ref{der}). Note that, due to the 
absence of external fields, the supplementary contribution 
$\Psi_\mu$ to ${\cal F}_\mu$, in Eq.~(\ref{n3.3}), 
 is negligible, being of order $G^2$ because 
$h_{\mu \nu} + 2\alpha\,\vphi = {\cal O}(G)$  and 
$(\ddot{z}_\mu - z_{\mu}^{\prime \prime}) = {\cal O}(G)$. 
[Both the infinite part and the finite part of $\Psi_\mu$ are 
${\cal O}(G^2)$.]

The expressions (\ref{rforce1})--(\ref{rforce3}) are linear (non-local) functions of the 
field derivatives. Following Dirac \cite{Dirac38} it is useful to decompose 
any field $A_{\rm ret} (x)$ in two parts:
\bea
\label{4.1}
&& A_{\rm ret} (x) = A_{\rm sym} (x) + A_{\rm reac} (x) \,,\\
&& A_{\rm sym} (x) \equiv \frac{1}{2} \, (A_{\rm ret} (x) + 
A_{\rm adv} (x)) \,, \\
&& A_{\rm reac} (x) \equiv \frac{1}{2} \, A_{\rm rad} (x) \equiv \frac{1}{2} \, 
(A_{\rm ret} (x) - A_{\rm adv} (x)) \,.
\eea
Note the definition of two fields, $A_{\rm reac}$ and $A_{\rm rad}$, differing 
by a factor 2, associated to the difference $A_{\rm ret} - A_{\rm adv}$. Both 
fields play a special role in the discussion below. They are both finite, as 
well as their derivatives, when considered at a point $x=z$ of the source. 
Therefore the contribution to the self-force corresponding to $A_{\rm reac}$ 
is finite and does not need to be renormalized. Hence, we shall dispense in 
the following with the label ${\rm R}$ when considering ${\cal F}_{\rm R}^{\mu} (A_{\rm 
reac})$. To simplify the notation we henceforth drop the label ``lin''
on ${\cal F}_\mu$, and freely move indices by $\eta_{\mu \nu}$ because 
we shall consistently work only to first order in $G$.
[As said above, in the present (first-order, no-external-field) approximation, 
we could even formally dispense with renormalizing ${\cal F}^\mu(A_{\rm sym})$ 
because the divergent contributions (\ref{3.20})--(\ref{3.22}) are ${\cal O}(G(
\ddot{z}_\mu -z^{\prime \prime}_\mu))={\cal O}(G^2)$. But, for clarity 
we continue to work with ${\cal F}_{\rm R}^\mu(A_{\rm sym})$.] 
\subsection{Reactive part of the self-force}

Let us first prove why, very generally, in the decomposition of the force 
corresponding to Eq. (\ref{4.1}),
\beq
{\cal F}_{\rm R}^{\mu} (A_{\rm ret}) = {\cal F}_{\rm R}^{\mu} (A_{\rm sym}) + {\cal 
F}^{\mu} (A_{\rm reac}) \equiv {\cal F}_{\rm R sym}^{\mu} + {\cal F}_{\rm 
reac}^{\mu}\,,
\eeq
the term ${\cal F}_{\rm reac}^{\mu}$ can be considered as defining the full 
radiation reaction force, responsible for draining out of the mechanical system 
on which it acts (the string in our case) the energy lost to infinity in the 
form of waves of the $A$ field. Indeed, for any field (in the linear 
approximation) we can define a field (pseudo) energy-momentum tensor $T_f^{\mu \nu} 
(A)$ which is quadratic in (the derivatives of) $A$. The total energy tensor 
$T^{\mu \nu} = T_s^{\mu \nu} + T_f^{\mu \nu}$ (where $T_s^{\mu \nu}$ denotes 
the energy tensor of source including any possible field-interaction 
energy localized on the source) is conserved: $0 = \partial_{\nu} \, T^{\mu 
\nu}$. This leads to the equations of the source: $\partial_{\nu} \, T_s^{\mu 
\nu} = F^{\mu} (A)$ where $F^{\mu} (A) \equiv - \partial_{\nu} \, T_f^{\mu 
\nu} (A)$ represents the spacetime (rather than worldsheet) version of the 
force density acting on the source. [We work here with the bare force 
density.] Let us consider, as a formal simplification, the case where the 
coupling between the source and the field $A$ is (adiabatically) turned off in 
the far past and the far future. [This means, in particular, that any possible 
field-interaction energy localized on the source vanishes in the far past and 
the far future.] Then the energy-momentum lost by the source 
during the entire interaction with the field, $P_{s \, {\rm lost}}^{\mu} = - 
\int d^3 x \, [T_s^{\mu 0} (+\infty) - T_s^{\mu 0} (-\infty)]$,
can be written 
as
\beq
P_{s \, {\rm lost}}^{\mu} = - \int d^4 x \, \partial_{\nu} \, T_s^{\mu \nu} = 
- \int d^4 x \, F^{\mu} (A) = P_{f \, {\rm gained}}^{\mu} \, ,
\eeq
where $P_{f \, {\rm gained}}^{\mu} = + \int d^3 x \, [T_f^{\mu 0} (+\infty) - 
T_f^{\mu 0} (-\infty)]$ is the energy-momentum gained by the field. When 
applying this result to the usual interaction force $F^{\mu} (A_{\rm ret}) = 
-\partial_{\nu} \, T^{\mu \nu}_f \, (A_{\rm ret})$ one has zero energy in 
$A_{\rm ret}$ in the far past, so that $P_{s \, {\rm lost}}^{\mu} = \int d^3 x 
\, T_f^{\mu \nu} (A_{\rm ret} (t=+\infty))$. The field energy momentum tensor 
$T_f^{\mu \nu} (A)$ is quadratic in the field and can always be written as the 
diagonal value of a symmetric quadratic form $T_f^{\mu \nu} (A) = Q^{\mu \nu} 
(A,A)$. It is easy to see that the generic structure $F^{\mu} (A) 
\equiv - \partial_{\nu} \, T_f^{\mu \nu} (A) = {\cal S}_A \cdot \partial \, 
A$, where ${\cal S}_A$ is a source term for the field $A$, and where the dot 
product denotes some contraction of indices, is generalized, when considering 
$Q^{\mu \nu}$ to: $- \partial_{\nu} \, Q^{\mu \nu} (A_1 , A_2) = \frac{1}{2} 
\, [{\cal S}_{A_1} \cdot \partial \, A_2 + {\cal S}_{A_2} \cdot \partial \, 
A_1]$. We can apply this to the case where $A_1 = A_{\rm ret}$ and $A_2 = 
A_{\rm rad} = A_{\rm ret} - A_{\rm adv}$ (for  which ${\cal S}_{A_2} = 0$) 
with the result:
\beq
\frac{1}{2} \, {\cal S} \cdot \partial \, A_{\rm rad} = {\cal S} \cdot 
\partial \, A_{\rm reac} = F^{\mu} (A_{\rm reac}) = - \partial_{\nu} \, Q^{\mu 
\nu} (A_{\rm ret} , A_{\rm rad})\,,
\eeq
where ${\cal S}$ is the usual source, and $F^{\mu} (A_{\rm reac})$ the result 
of replacing $A_{\rm ret}$ by $A_{\rm reac} = \frac{1}{2} \, A_{\rm rad}$ in 
the usual force density. Integrating the latter formula over spacetime gives
\beq
-\int d^4 x \, F^{\mu} (A_{\rm reac}) = \int d^3 x \, [Q^{\mu 0} (A_{\rm ret} 
, A_{\rm rad})_{{\big |}_{t=+\infty}} - Q^{\mu 0} (A_{\rm ret} , A_{\rm rad})_ 
{{\big |}_{t=-\infty}}] \, .
\eeq
Again, one has zero energy from the far past contribution (because $A_{\rm 
ret} (-\infty) = 0$), while the far future contribution is simply, thanks to 
$A_{\rm adv} (+\infty) = 0$, $Q^{\mu 0} (A_{\rm ret} , A_{\rm ret}) = T_f^{\mu 0}
(A_{\rm ret})$ so that
\beq
-\int d^4 x \, F^{\mu} (A_{\rm reac}) = \int d^3 x \, T_f^{\mu 0} (A_{\rm ret} 
(+\infty)) = -\int d^4 x \, F^{\mu} (A_{\rm ret}) = P_{s \, {\rm lost}}^{\mu} 
\, .
\eeq
This proves, for any field treated in the linear approximation, that the 
contribution to the self-force due to $A_{\rm reac}$ contains, when integrated 
over time, the full effect of radiation damping, ensuring conservation with the 
energy-momentum lost to radiation. The contribution $F^{\mu} (A_{\rm reac})$ 
can be called the ``reactive'' part of the self-force $F^{\mu} (A_{\rm ret})$. 

Summarizing the results at this point, the renormalized self-interaction force 
(returning now to the worldsheet distributed force density) can be written as
\beq
{\cal F}_{\rm R}^{\mu} = \mbox{FP} {\cal F}^{\mu} = 
\mbox{FP}{\cal F}_{\rm sym}^{\mu} (\delta) + {\cal F}_{\rm reac}^{\mu} \,,
\label{4.9}
\eeq
where $\mbox{FP}$ denotes Hadamard's Finite Part (``Partie Finie'') 
operation \cite{Had} (i.e., in our case, the result of subtracting a term 
$\propto \log(\delta / \Delta_{\rm R})$ from the ultraviolet-cutoff integral ${\cal 
F}^{\mu} (\delta) = \int_{\sigma_0}^{\sigma - \delta_c} d\sigma' [\ldots] + 
\int_{\sigma + \delta_c}^{\sigma_0 +L} d\sigma' [\ldots]$). Note that only the 
{\it symmetric}\/ contribution, obtained by replacing $A_{\rm ret}$ by $A_{\rm 
sym} = \frac{1}{2} \, (A_{\rm ret} + A_{\rm adv})$ in the force density, needs 
to be renormalized (and, as we said above, one can even formally 
dispense with considering this renormalization).
This symmetric contribution does not contribute, after integration 
over time, to the overall damping of the source. The finite  
reactive contribution ${\cal F}_{\rm reac}^{\mu} 
\equiv {\cal F}^{\mu} (A_{\rm reac})$ embodies (on the average) the full 
effect of radiation damping.

The advantage of the above decomposition is to isolate, very cleanly, the 
radiation damping force from the other non-cumulative, self-interactions. Its 
disadvantage is to write the non-local, but {\it causal}\/ self-force 
$\mbox{FP}\,{\cal F}^{\mu} (A_{\rm ret})$ 
as a sum of two {\it acausal}\/ (meaning 
future-dependent) contributions. Indeed, both $\mbox{FP}({\cal F}_{\rm sym}^\mu)$ 
and ${\cal F}_{\rm reac}^\mu$ are given by integrals whose support is the 
intersection of the worldsheet with the {\it two-sided} light 
cone with vertex located at $z^\mu$. 
In principle one can work directly with the 
full, causal ${\cal F}_{\rm R}^{\mu}$ (as done, e.g., in Ref. 
\cite{QuashnockSpergel90}), but this is computationally very intensive. [A 
simplification, used by the latter authors, and mentioned above, 
is that the self-force ${\cal 
F}^{\mu} (\delta)$ becomes, as is clear from Eqs. (\ref{3.20})--(\ref{3.22}), 
finite as $\delta 
\rightarrow 0$ when evaluated on free-string trajectories, satisfying ${\ddot 
z}^{\mu} - z^{\mu\,\prime \prime} = 0$.] We shall follow Refs. \cite{BS95}, \cite{BS96} in 
working only with the (finite) reactive force ${\cal F}_{\rm reac}^{\mu}$ and 
in trying to define a simple local approximation for it.

\subsection{Local back-reaction terms in dimensional regularization}
\label{subsec4.2}
The reaction force ${\cal F}_{\rm reac}^{\mu}$ is linear in $\partial_{\mu} \, 
A_{\rm reac} (z)$, which is itself given by the following integral
\beq
\label{4.10}
\partial_{\mu} \, A_{\rm reac} (z) = \int_0^L d\sigma' \, B_{\mu}^z (\sigma') 
\, ,
\eeq
with
\beq
B_{\mu}^z (\sigma') = \frac{1}{2} \, \Biggl\{ \left[ \frac{1}{\vert \Omega 
\cdot \dot z \vert} \, \frac{d}{d \tau'} \, \left( 
\frac{\Omega_{\mu} \, \Sigma (\sigma' , \tau')}{\Omega \cdot \dot z} \right) 
\right]_{{\big |}_{{\tau' = \tau_{\rm ret}}}}
- \left[ \frac{1}{\vert \Omega \cdot \dot z \vert} \, 
\frac{d}{d \tau'} \, \left( \frac{\Omega_{\mu} \, \Sigma 
(\sigma' 
, \tau')}{\Omega \cdot \dot z} \right) \right]_{{\big |}_{{\tau' = \tau_{\rm adv}}}}
\Biggl\}\, . 
\eeq
The integrand $B_{\mu}^z (\sigma')$ is the finite difference between two terms 
that blow up when $\sigma' \rightarrow \sigma$ ($\sigma$ being such that 
$z=z(\sigma , \tau)$). When $\sigma'$ is well away from $\sigma$ (say, for 
long, horizon-sized strings) $B_{\mu}^z (\sigma')$ is expected to decrease 
roughly as the inverse spatial distance $\vert \Omega \cdot \dot z \vert$, 
i.e. 
roughly as $\vert \sigma' - \sigma \vert^{-1}$. In other words, a very rough 
representation of the typical behaviour of $B_{\mu} (\sigma)$ is $B(\sigma') 
\sim (2(\sigmap-\sigma))^{-1} \, [f(\tau - (\sigmap-\sigma)) - f(\tau + (\sigmap- 
\sigma)]$, where the ``effective source function'' $f(\tau)$ is expected to 
oscillate as $\tau$ varies. If we 
think in terms of one Fourier mode, say $f(\tau) = f_{\omega} \, e^{-i \omega 
\tau}$, these considerations suggest that the field derivative $\partial \, 
A$ is roughly given by an integral of the form
\beq
\partial \, A = \int d\sigma' \, B(\sigma') \sim i f_{\omega} \, e^{-i \omega 
\tau} \int_{-\infty}^{+\infty} d\sigma' \,\frac{\sin \omega (\sigmap-\sigma)}
{(\sigmap-\sigma)} \, .
\eeq
The latter integral is equal to $\pi$, so that one can finally replace the 
oscillatory and decreasing integrand $B(\sigma')$ by an effective 
$\delta$-function, $B_{\rm eff} (\sigma') = B(0) \, \Delta \,\delta (\sigma' - 
\sigma)$, with (in our example) $B(0) = -\dot f (0) = i\omega f_{\omega}$ and 
$\Delta = \pi / \omega$, or, in other words, $\partial \, A = \int d\sigma' \, 
B(\sigma')$ is replaced by $\Delta \, B(0)$. The analogous proposal of 
replacing the complicated, non-local integral (\ref{4.10}) giving
$\partial_{\mu} \, A_{\rm reac}$ simply by the local expression
\beq
\label{locapp}
[\partial_{\mu} \, A_{\rm reac}]^{\rm local} = \Delta \, B_{\mu}^z (0) \, ,
\eeq
where $\Delta$ is some length scale linked to the wavelength of the main 
Fourier component of the radiation, was made by Battye and Shellard 
\cite{BS95}, \cite{BS96} (see also \cite{DQ90}). In effect, this proposal is 
equivalent to replacing the $\sigma^\prime$-extended source $\Sigma(\taup,\sigmap)$ 
by the $\sigma^\prime$-local effective source $\Delta\,\Sigma(\taup,\sigma)\,
\delta(\sigmap-\sigma)$. One of the main aims of the 
present paper is to study critically the consequences of this proposal.

Though this ``local back reaction approximation'' drastically simplifies the 
evaluation of the reaction force ${\cal F}_{\rm reac}^{\mu}$, there remains 
the non-trivial analytical task of computing the $\sigma' \rightarrow 0$ limit 
of the difference between the two complicated (and divergent) terms making up 
$B_{\mu}^z (\sigma')$. We found very helpful in this respect to use 
{\it dimensional regularization}, i.e. to use, instead of the normal (singular) four 
dimensional Green's functions (\ref{gret}), (\ref{gadv}), 
their analytic continuation to a spacetime of (formal) 
dimension $n = 4-\epsilon$. [We shall keep computing 
the index algebra in $4$-dimensions. This is allowed here because 
our use of dimensional regularization is, simply, a technical 
trick for computing the finite object $B_{\mu}^z(\sigma)$.] 
This technique is well known 
to be quite useful in quantum field theory, but it (or, at least, a variant of 
it) has also been shown long ago to be technically very convenient in the 
classical theory of point particles \cite{Riesz}, \cite{Fremberg}, \cite{Ma}, \cite{Damour75}.

Riesz \cite{Riesz} has shown that the retarded and advanced Green's functions 
in dimension $n \equiv 4-\epsilon$ read
\beq
\label{grie}
G_{{\rm ret} \atop {\rm adv}}^{(n)} (x-y) = \frac{1}{H_n (2)}\, 
(-(x-y)^2)^{\frac{2-n}{2}} \, \theta (-(x-y)^2) \, 
\theta (\pm (x^0 - y^0)) \,,
\eeq
with $H_n (2) = 2 \, \pi^{\frac{n-2}{2}} \, 
\Gamma \left( \frac{4-n}{2} \right)$ and 
\beq
\Box \, G^{(n)} (x-y) = -\delta^n (x-y) \, .
\eeq
Note that, when $\epsilon = 4-n \rightarrow 0$, the coefficient appearing in 
Eq. (\ref{grie}) becomes
\beq
\label{hn}
\frac{1}{H_n (2)} = \frac{\epsilon}{4\pi}\, (1+{\cal O} (\epsilon)) \, .
\eeq
To save writing, we shall neglect in the following the factor $1+{\cal O} 
(\epsilon)$ in Eq. (\ref{hn})
which plays no role in the terms we consider. Then, we 
write the retarded solution of our model field equation (2.1) in dimension 
$n=4-\epsilon$ as
\beq
\label{ret1}
A_{\rm ret}(x) =\epsilon\,\int
 d \sigmap \int_{-\infty}^{\tau_{\rm ret}}
d \taup \,\Sigma\,(\Omega^2)^{\frac{2-n}{2}}\,\theta(\Omega^2) \,,
\eeq
where $\Omega^2 \equiv -(x-z(\taup,\sigmap))^2$. (Note the inclusion of a minus sign so that 
$\Omega^2 > 0$ within the light cone). Again neglecting a factor $1 + {\cal O} 
(\epsilon)$, the field derivative reads
\beq
\label{dret1}
\pa_\mu A_{\rm ret}(x) = 2\epsilon\,\int
 d \sigmap \int_{-\infty}^{\tau_{\rm ret}}
d \taup \,\Sigma\,\Omega_\mu\,(\Omega^2)^{-{n}/{2}}\theta(\Omega^2) \,.
\eeq
Using some efficient tools of dimensional regularization 
(which are explained in Appendix A) we get our main technical results:
the explicit expressions of the reactive field, and its derivatives, in the local 
back reaction approximation
\bea 
\label{4.32}
{[A_{\rm reac}(z)]}^{\rm local} &=& \frac{\Delta}{\dot{z}^2}\,\left [
\dot{\Sigma} - \Sigma\,\left (\frac{\dot{z} \cdot 
\ddot{z}}{\dot{z}^2} \right )\right] \,,\\
\label{4.33}
{[\pa_\mu A_{\rm reac}(z)]}^{\rm local} &=& 
\frac{\Delta}{(\dot{z}^2)^2}\,\left [ 
\frac{1}{3}\Sigma\,\tdot{z}_\mu + 
\dot{\Sigma}\,\ddot{z}_\mu + \ddot{\Sigma}\,\dot{z}_\mu - 
4\dot{\Sigma}\,\dot{z}_\mu \, 
\left ( \frac{\dot{z}\cdot \ddot{z}}{\dot{z}^2} \right )
-2 \Sigma \ddot{z}_\mu\, \left ( \frac{\dot{z}\cdot \ddot{z}}{\dot{z}^2} 
\right )\right .\nonumber \\
&& \left . -\Sigma\,\dot{z}_\mu \,\left ( \frac{\ddot{z}\cdot 
\ddot{z}}{\dot{z}^2} \right )
- \frac{4}{3}\,\Sigma\,\dot{z}_\mu\,\left ( \frac{\dot{z} \cdot 
\tdot{z}}{\dot{z}^2} \right )
+ 6 \Sigma\,\dot{z}_\mu\,\left ( \frac{\dot{z}\cdot \ddot{z}}{\dot{z}^2} \right )^2
\right ]\,.
\eea
Some simplifications occur if we introduce, instead of the worldsheet density 
$\Sigma$, the corresponding worldsheet scalar $S \equiv \Sigma / 
\sqrt{\gamma}$. 
We find
\bea
\label{4.36}
{[A_{\rm reac} (z)]}^{\rm local} &=& \Delta \, \left [-\dot S -
S\,\left (\frac{\dot{z} \cdot \ddot{z}}{\dot{z}^2} \right )\right ]\,, \\
\label{4.37}
{[\partial_{\mu} \, A_{\rm reac} (z)]}^{\rm local} &=& 
\frac{\Delta}{\dot{z}^2}\, \left \{ S\,\left[ 
-\frac{1}{3}\,{\tdot z}_{\mu} -
\frac{2}{3}\,\dot{z}_\mu\,\left (
\frac{\dot{z}\cdot \tdot{z}}{\dot{z}^2}\right ) + 
2 \dot{z}_\mu\,\left (\frac{\dot{z}\cdot \ddot{z}}{\dot{z}^2} \right )^2
-\dot{z}_\mu\,\left (\frac{\ddot{z}\cdot \ddot{z}}{\dot{z}^2} \right )
\right ] \right . \nonumber \\
&& \left . -\ddot{z}_\mu\,\dot{S} - \dot{z}_\mu\,\ddot{S} \right \} \,.
\eea
Note that Eqs.~(\ref{4.32}), (\ref{4.33}) and 
Eqs.~(\ref{4.36}), (\ref{4.37}) satisfy the compatibility condition 
${\dot z}^{\mu} \, \partial_{\mu} \, A 
= \dot A$, but because of the lack of worldsheet covariance (broken by 
the introduction of $\Delta$) the analog condition 
for $z'$ is not verified.

\subsection{Dilaton radiation reaction}
\label{subsec4.3}
Let us first apply our results to the case of the dilaton field $\varphi$, which 
has not been previously studied in the literature. The corresponding 
(worldsheet scalar) source is then simply
\beq
S_{\varphi} = \frac{1}{z'^2} \, \Sigma_{\varphi} = \frac{1}{z'^2} \, \alpha \, 
G {\mu} \, U = -2 \, \alpha \, G {\mu} \, .
\eeq
$S_{\varphi}$ being a constant, the preceding formulas simplify very much:
\bea
&& [\varphi_{\rm reac} (z)]^{\rm local} = 2\alpha\,G\,\mu\,\Delta 
\, \left( \frac{\dot{z}\cdot \ddot{z}}{\dot{z}^2} \right ) \,,\\
&& [\partial_{\mu} \, \varphi_{\rm reac} (z)]^{\rm local} =
2\alpha\,G\,\mu\,\frac{\Delta}{\dot{z}^2}\,\left[ 
\frac{1}{3}\,{\tdot z}_{\mu} +
\frac{2}{3}\,\dot{z}_\mu\,\left (
\frac{\dot{z}\cdot \tdot{z}}{\dot{z}^2}\right ) - 
2 \dot{z}_\mu\,\left (\frac{\dot{z}\cdot \ddot{z}}{\dot{z}^2} 
\right )^2 + \dot{z}_\mu\,\left (\frac{\ddot{z}\cdot \ddot{z}}{\dot{z}^2}\right )
\right ] \,.
\eea
Inserting these results in the dilaton self-force (\ref{rforce1}), we get
\beq
\label{reacd}
{\cal F}_\mu^{\vphi\,\rm local} = 
\frac{4}{3}\alpha^2\,G\,\mu^2\,\Delta\,\left [ 
\tdot{z}_\mu - \dot{z}_\mu\,\left ( \frac{\dot{z}\cdot \tdot{z}}{\dot{z}^2} \right ) 
+ {z}^{\prime}_{\mu}\,\left ( \frac{{z}^\prime 
\cdot \tdot{z}}{\dot{z}^2} \right )\right ]\,.
\eeq
For notational simplicity, we henceforth drop the label ``local''
on the local approximations to the reactive forces.
Consistently with our choice of conformal gauge (which, in the case of the 
dilaton coupling, is the same as in flat space, see Eq. (\ref{conf})), we see that the 
reaction force (\ref{reacd}) is orthogonal to the two 
worldsheet tangent vectors, ${\dot z}^{\mu}$ and $z'^{\mu}$:
\beq
{\dot z}^{\mu} \, {\cal F}_{\mu}^{\varphi} \equiv 0 \equiv z'^{\mu} \, {\cal 
F}_{\mu}^{\varphi} \, .
\eeq
Let us now show that the putative, local approximation to the dilaton reaction 
force, Eq. (\ref{reacd}) conveys some of the correct physical characteristics expected 
from a radiation damping force. In particular, let us check that the overall 
sign of Eq. (\ref{reacd}) is the correct one. First, we remark that we can work 
iteratively and therefore consider that the reaction force (\ref{reacd}), and its 
integrated effects, can be evaluated on a free string trajectory. In other 
words, when evaluating the total four momentum lost, 
$P_s^{\rm lost} = P^{\rm lin}_{s \mu}(-\infty) - P^{\rm lin}_{s \mu}(+\infty)$ 
(with, say, the convenient definition 
$P^{\rm lin}_{s \mu}(\tau) \equiv \int_0^L d \sigma\,\mu_{\rm R}\,\dot{z}_\mu(\sigma,\tau)$), 
by the string under the action of 
${\cal F}_{\mu} = {\cal F}_{\mu}^{\varphi} + {\cal F}_{\mu}^h + {\cal 
F}_{\mu}^B$,
\beq
\label{momd}
P_{s \, \mu}^{\rm lost} = -\int d\sigma \, d\tau \, {\cal F}_{\mu} \, ,
\eeq
we can insert a free string trajectory on the right-hand side of 
(\ref{momd})\footnote[1]{Strictly speaking the integral in Eq. (\ref{momd}) is infinite 
because free string trajectories are periodic. The meaning of Eq. (\ref{momd}), and 
similar integrals below is to give, after division by the total coordinate time 
span $\tau$, the time-averaged energy-momentum loss.}. This being the case, we 
can now further restrict the worldsheet gauge by choosing a {\it temporal} 
conformal gauge, i.e. such that $t=z^0 (\tau , \sigma) = \tau$. Geometrically, 
this means that the $\tau ={\rm const.}$ sections of the worldsheet coincide with 
$x^0 = {\rm const.}$ space-time coordinate planes. [The choice $z^0 = \tau$ is 
consistent for free string trajectories because for them ${\ddot z}^{\mu} - 
z^{\mu\,\prime \prime} = 0$.] In this gauge, we have
\beq
\dot{z}^0 =1\,,\quad \quad -\dot{z}^2 = 1 - {\bf v}^{\,2} = {\bf z}^{\,\prime\,2}\,,
\quad \quad \dot{z}\cdot \tdot{z} = {\bf v}\cdot {\ddot{\bf v}} \, \quad \quad 
{z}^\prime \cdot \tdot{z} = {{\bf z}}^{\,\prime} \cdot {\ddot{\bf v}} \,,
\eeq 
where we have introduced the 3-velocity ${\bf v} \equiv \dot{\bf z}$. The zero 
component of Eq.~(\ref{momd}) then reads
\beq
{\cal F}_{\varphi}^0 = + \frac{4}{3} \, \alpha^2 \, G \mu^2 \, \Delta \, 
\frac{{\bf v} \cdot \ddot{\bf v}}{1- {\bf v}^2} \, .
\eeq
Assuming that the scale $\Delta$ is constant, we can integrate by parts and 
write for the total energy lost by the string
\beq
\label{ened}
E_{\varphi}^{\rm lost} = \frac{4}{3}\alpha^2\,G\,\mu^2\,\Delta\,
\int d \sigma \,d \tau 
\, \left [\frac{\dot{\bf v}^2}{1 -{\bf v}^2}+ 
2\,\frac{ ({{\bf v}}\cdot \dot{{\bf v}})^2}{(1 -{\bf v}^2)^2}
\right ]\,.
\eeq
The integrand of Eq. (\ref{ened}) is positive definite, ensuring that the reaction 
force (\ref{reacd}) has the correct sign for representing a radiation damping force.

We can further check that the total 4-momentum lost by the string is, as it 
should, time-like. First, let us note that the relation
\beq
U^{\rho \mu}\,\pa_\rho \vphi = -\pa_a(\sqrt{\gamma}\,\gamma^{a b}\,\pa_b z^\mu 
\, \vphi) + \vphi\,\pa_a(\sqrt{\gamma}\,\gamma^{a b}\,\pa_b z^\mu) \,,
\eeq
shows that, as far as its integrated effects are concerned, the dilaton reaction 
force (\ref{f1}) is equivalent to
\beq
\label{p2}
{\cal F}_{\mu}^{\varphi \, {\rm equiv.}} = \alpha \, \mu \, U \, 
\partial_{\mu} \, \varphi_{\rm reac} = \frac{1}{G} \, \Sigma_{\varphi} \, 
\partial_{\mu} \, \varphi_{\rm reac} \, .
\eeq
Inserting Eq. (\ref{4.33}), or better, Eq. (\ref{4.37}) 
into Eq. (\ref{p2}) yields, after 
integration by parts, a total 4-momentum loss
\beq
\label{4.49}
P_{\mu}^{{\rm lost} \, \varphi} = \frac{4}{3} \, \alpha^2 \, G \mu^2 \, \Delta 
\int d\tau \, d\sigma \, \pi_{\mu} \, ,
\eeq
with integrand
\beq
\label{p5}
\pi_{\mu} = \dot{z}_\mu\,\left [ 2\left (\frac{\dot{z} \cdot \ddot{z}}{\dot{z}^2} \right )^2- \left ( \frac{\ddot{z} \cdot \ddot{z}}{\dot{z}^2} \right ) \right ]
+ 2\ddot{z}_\mu\,\left ( \frac{\dot{z} \cdot \ddot{z}}{\dot{z}^2} \right )\,.
\eeq
The square of $\pi_{\mu}$ reads
\beq
\pi_{\mu} \, \pi^{\mu} = \frac{1}{(\dot{z}^2)^4}\, \left [
12\dot{z}^2\,(\dot{z}\cdot \ddot{z})^4 + (\dot{z}^2)^3\,(\ddot{z}^2)^2 - 
4\ddot{z}^2\,(\dot{z}^2)^2\,(\dot{z}\cdot \ddot{z})^2 \right ]\,.
\eeq
This is negative definite in the temporal gauge $t=\tau$, showing that $\vert 
{\bf P}_{\varphi}^{\rm lost} \vert < E_{\varphi}^{\rm lost}$, as physically 
expected.

\subsection{Gravitational and axionic radiation reaction}
\label{subsec4.4}
We are going to see that the generalization of the dilaton results to the case 
of the gravitational and axionic fields is non-trivial, and leads to physically 
nonsensical results. Let us first generalize Eq.~(\ref{p2}). The relations
\bea
&& U^{\alpha \beta}\,\pa_\alpha h_{\beta \mu} = 
-\pa_a(\sqrt{\gamma}\,\gamma^{a b}\,h_{\alpha \mu}\,\pa_b z^\alpha) +
h_{\alpha \mu}\,\pa_a(\sqrt{\gamma}\,\gamma^{a b}\,\pa_b z^\alpha) \,, \\
&& V^{\lambda \nu}\,\pa_\lambda B_{\nu \mu} = 
\pa_b (\epsilon^{a b}\,\pa_a z^\nu\,B_{\nu \mu}) - 
\pa_b (\epsilon^{a b}\,\pa_a z^\nu)\,B_{\nu \mu}\,, 
\eea
show that, as far as their integrated effects are concerned, the gravitational 
and axionic reaction forces (\ref{f2}), (\ref{f3}) are equivalent, respectively, to:
\bea
&& {\cal F}_{\mu}^{h \, {\rm equiv.}} = \frac{1}{2} \, \mu \, U^{\alpha \beta} 
\, \partial_{\mu} \, h_{\alpha \beta}^{\rm reac} \, ,\\
&& {\cal F}_{\mu}^{B \, {\rm equiv.}} = \frac{1}{2} \, \lambda \, V^{\alpha 
\beta} \, \partial_{\mu} \, B_{\alpha \beta}^{\rm reac} \, .
\eea
It is important to note that, as in the dilaton case Eq. (\ref{p2}), these equivalent 
reaction forces are simple bilinear forms in the vertex operators and the 
derivatives of the fields. They can both be written as
\beq
\label{p3}
{\cal F}_{\mu}^{\rm equiv.} = \frac{1}{8G} \, \Sigma \cdot \partial_{\mu} 
\, A_{\rm reac}
\eeq
where, as in Eq. (\ref{box}), $\Sigma$ denotes the source of the field $A = h_{\alpha 
\beta}$ or $B_{\alpha \beta}$, and where the dot denotes a certain symmetric 
bilinear form acting on symmetric or antisymmetric tensors. With the 
normalization of Eq. (\ref{p3}) these bilinear forms are, respectively,
\bea
\label{4.57}
&& U_{\alpha \beta} = U_{\beta \alpha}: \,\,\,\quad \quad U \cdot U \equiv U_{\alpha \beta} \, 
U^{\alpha \beta} - \frac{1}{2} \, U^2 \equiv U_{\alpha \beta} \, 
\widetilde{U}^{\alpha \beta} \,, \\ 
\label{4.58}
&& V_{\alpha \beta} = -V_{\beta \alpha}: \quad \quad 
V \cdot V \equiv V_{\alpha \beta} \, V^{\alpha \beta} \,.
\eea
One can recognize here the quadratic forms defined by the residues of the 
gauge-fixed propagators of the $h$ and $B$ fields. 
Note that if we wish to rewrite the scalar reaction force (\ref{p2}) 
in the same format (\ref{p3}) as the tensor ones we have to define 
the dot product for scalar sources as  
\beq
\label{new}
\Sigma_\vphi \cdot \Sigma_\vphi \equiv 8\,\Sigma^2_\vphi\,.
\eeq
Using this notation, and the 
results above on the reaction fields, it is possible to compute in a rather 
streamlined way the total 4-momentum lost under the action of the local 
reaction force:
\beq
P_{\mu}^{\rm lost} = -\frac{1}{8G} \int \int d\sigma \, d\tau \, \Sigma \cdot 
[\partial_{\mu} \, A_{\rm reac}]^{\rm local} \, .
\eeq
The calculation is simple if one uses the form (\ref{4.37}). Let us note that the 
worldsheet-scalar sources $(S = \Sigma / \sqrt{\gamma})$ for the three fields 
we consider ($\varphi$, $h$ and $B$) satisfy
\bea
\label{ss1}
&& S \cdot S = \hbox{const.} \,,\\
\label{ss2}
&& S \cdot \dot S = 0 \, .
\eea
Indeed, if we introduce the scalarized vertex operators (with conformal 
dimension zero) $\widehat U \equiv U / \sqrt{\gamma}$, $\widehat{U}_{\alpha 
\beta} \equiv U_{\alpha \beta} / \sqrt{\gamma}$ and $\widehat{V}_{\alpha \beta} 
\equiv V_{\alpha \beta} / \sqrt{\gamma}$, it is easily seen that
\beq
\label{uu}
\matrix{
\varphi : & \widehat{U}\cdot \widehat{U} \equiv 8\,\widehat{U}^2 
= +32 \, , \cr
h: &\widehat{U}_{\alpha \beta} \cdot \widehat{U}^{\alpha \beta} \equiv 
\widehat{U}_{\alpha \beta} \, \widehat{U}^{\alpha \beta} - \frac{1}{2} \, 
\widehat{U}^2 = 0 \, , \cr
B: &\widehat{V}_{\alpha \beta} \cdot \widehat{V}^{\alpha \beta} = -2 \, . \cr
}
\eeq
The relations 
(\ref{ss1}), (\ref{ss2}) simplify very much the evaluation of $P_{\mu}^{\rm 
lost}$. In particular, the constancy of $S \cdot S$ allows one to integrate by 
parts on ${\tdot z}_{\mu}$, etc $\ldots$ without having to differentiate the $S 
\cdot S$ factors. By some simple manipulations, using also the consequence
\beq
\dot S \cdot \dot S + S \cdot \ddot S = 0 \, ,
\eeq
of Eq. (\ref{ss2}), we get
\beq
\label{ris1}
P_{\mu}^{\rm lost} = \frac{1}{8G} \int \int d\sigma \, d\tau \int {\dot 
z}_{\mu} \, \left \{ ({\dot S}\cdot {\dot S}) + (S\cdot S) 
\, \left [ \frac{1}{3} \, {\dot u}^2 + 
\frac{1}{12} \, {\dot \phi}^2 \right ] + \frac{1}{3} \, {\ddot z}_{\mu} 
\, (S \cdot S) \, \dot \phi \right \} \, .
\eeq
Here we introduced a special notation for the conformal 
factor (Liouville field),  
\beq
ds^2 = e^{\phi} (-d\tau^2 + d\sigma^2) \,, \quad \quad  
e^{\phi} = \sqrt{\gamma} = z'^2 = -{\dot z}^2 \,,
\eeq
and we defined the {\it unit} time like vector $u^\mu = e^{-\phi/2}\,\dot{z}^\mu$, 
and its first derivative
\beq
\dot{u}_\mu = \frac{d}{d \tau}\, (e^{-\phi/2}\,\dot{z}_\mu)\,, 
\quad \quad \dot{u}^2 = - \frac{\ddot{z}^2}{\dot{z}^2} + \frac{1}{4}\,\dot{\phi}^2
> 0\,.
\eeq
Let us now prove the remarkable result that the 
contribution proportional to ${\dot S}\cdot {\dot S}$ 
in Eq.~(\ref{ris1}) vanishes for all three 
fields when evaluated (as we are iteratively allowed to do) on a free string 
trajectory:
\beq
\label{p4}
\int \int d\sigma \, d\tau \, {\dot z}_{\mu} \, ({\dot S}\cdot {\dot S}) = 0 \, .
\eeq
Indeed, for the scalar case $\widehat U = -2$ and $\dot{\widehat U} = 0$, while 
for the other fields a straightforward calculation gives
\bea
&& {\dot{\widehat U}}_{\alpha \beta} \, {\dot{\widehat U}}^{\alpha \beta} \, - 
\frac{1}{2} \,  {\dot{\widehat U}}^2 = - \Box_{\eta} \, \phi \, ,\\
&& {\dot{\widehat V}}_{\alpha \beta} \, {\dot{\widehat V}}^{\alpha \beta} = + 
\Box_{\eta} \, \phi \, ,
\eea
when taking into account the vanishing of terms proportional to the worldsheet 
derivatives of $\Box_{\eta} \, z^{\mu} = -{\ddot z}^{\mu} + z^{\mu\,\prime 
\prime}$. [These results have a nice geometrical interpretation 
linked to the Gauss-Codazzi relations.] 
Integrating by parts, we see that the contribution (\ref{p4}) is 
proportional to $\int \int d\sigma \, d\tau \, (\Box_{\eta} \, {\dot z}_{\mu}) 
\, \phi$ which vanishes, again because of the free string equations of motion.

Finally, remembering the constancy of $S \cdot S$, we get the very 
simple result
\beq
P_{\mu}^{\rm lost} = \frac{1}{3} \, \Delta \, \frac{S \cdot S}{8G} \int \int 
d\sigma \, d\tau \, \pi_{\mu} \, ,
\eeq
where the integrand
\beq
\pi_{\mu} = {\dot z}_{\mu} \, \left({\dot u}^2 + \frac{1}{4} \, {\dot \phi}^2 
\right) + {\ddot z}_{\mu} \, \dot \phi
\eeq
is easily seen to coincide with the one which appeared above, Eq. (\ref{p5}), in 
our direct calculation of the dilaton reaction. Let us recall that the present 
calculation applies uniformly 
to all three fields if we define the dot product between dilatonic 
vertex operators with an extra factor $8$, see Eq. (\ref{new}).

The conclusion is that the local approximation to back reaction for the three 
fields $\varphi$, $h$ and $B$ leads to energy-momentum losses which are 
proportional to the same quantity $\int \int d^2 \, \sigma \, \pi_{\mu}$ with 
coefficients respectively given by (using Eqs. (\ref{uu}) above)
\bea
\label{n7}
&& \frac{\Delta}{3} \, \frac{S_{\varphi} \cdot S_{\varphi}}{8G} =  
\frac{\Delta}{3} \, (\alpha G \mu)^2 \, 
(\widehat U )^2 = + \frac{4}{3} \, \Delta \, G \, \alpha^2 \, \mu^2 \,, \\
\vrbig
\label{n8}
&& \frac{\Delta}{3} \, \frac{S_h \cdot S_h}{8G} = \frac{\Delta}{3} \, \frac{(4 G 
\mu)^2}{8G} \, \widehat{U}_{\alpha \beta} \, \widehat{\widetilde U}^{\alpha 
\beta} = 0 \,, \\ 
\vrsmall
&& \label{n9}
\frac{\Delta}{3} \, \frac{S_B \cdot S_B}{8G} = \frac{\Delta}{3} \, \frac{(4 G 
\lambda)^2}{8G} \, \widehat{V}_{\alpha \beta} \, \widehat{V}^{\alpha \beta} = - 
\frac{4}{3} \, \Delta \, G \, \lambda^2 \, .
\eea
The result (\ref{n7}) coincides with Eq. (\ref{4.49}) above (for which we have verified that the 
overall sign is correct). We therefore conclude that the ``local reaction 
approximation'' (\ref{locapp})  yields: (i) a {\it vanishing}, net energy-momentum loss for 
the gravitational field, and (ii) the {\it wrong sign} (antidamping) for the 
axionic field. The latter result disagrees with Refs. \cite{BS95}, \cite{BS96}
(see the Appendix) 
which claimed to obtain positive damping. It is for clarifying this important sign question 
that we have presented above a streamlined 
calculation showing that the overall sign can simply be read from the 
contraction of the vertex operators of the fields. Indeed, finally the physical 
energy-loss sign is simply determined by the easily checked (and {\it signature 
independent}) signs in Eqs. (\ref{uu}).

\subsection{Gauge invariance and mass-shell-only positivity}
\label{subsec4.5}
Why is the ``local back reaction approximation'' giving 
physically unacceptable answers in the 
cases of gravitational and axionic fields but a physically acceptable one in 
the case of the dilatonic field?
The basic reason for this difference between $h_{\mu \nu}$ and $B_{\mu \nu}$ on 
one side, and $\varphi$ on the other is the {\it gauge invariance} of the 
former. Indeed, a gauge symmetry (here $h_{\mu \nu} \rightarrow h_{\mu \nu} + 
\partial_{\mu} \, \xi_{\nu} + \partial_{\nu} \, \xi_{\mu}$, $B_{\mu \nu} 
\rightarrow B_{\mu \nu} + \partial_{\mu} \, A_{\nu} - \partial_{\nu} \, 
A_{\mu}$) means that some of the components of $h_{\mu \nu}$ and $B_{\mu \nu}$ 
are not real physical excitations. This is associated with the fact that some 
of the components of $h_{\mu \nu}$ and $B_{\mu \nu}$ (namely 
$h_{0i}$ and $B_{0i}$) 
have kinetic terms with the {\it wrong sign}, i.e. that they (formally) carry 
negative energy. Therefore, approximating radiation damping is very delicate 
for gauge fields. A slight violation of gauge invariance by the approximation 
procedure can lead to antidamping (the literature of gravitational radiation 
damping is full of such errors, see e.g. \cite{Damour87}). A more precise way 
of seeing why the local back reaction approximation is dangerous in this 
respect is the following.

We have proven above that an exact expression for the 4-momentum of the source 
lost to radiation is given (for $\varphi$, $h_{\mu \nu}$ and $B_{\mu \nu}$, and 
more generally for any linearly coupled field) by an expression of the form
\beq
\label{f}
P_{\mu}^{\rm lost} = -\,k \int d^4 x \, J(x) \cdot \partial_{\mu} \, A_{\rm reac} 
(x)\,,
\eeq
where $J(x)$ is the source of $A(x)$
\beq
\label{j}
\Box \, A(x) = -J(x) \, ,
\eeq
and where $k$ is a {\it positive} coefficient which depends on the normalization of 
the kinetic terms of $A(x)$ [$4\pi k = 1/8 G$ when using the above 
normalizations, the extra factor $4\pi$ compensating for our present way of 
writing the field equation $(\ref{j})$.] The spacetime source $J(x)$ is linked to our 
previous string distributed sources by $J(x) = 4\pi \int d^2 \sigma \, \Sigma 
\, \delta^4 (x-z)$. The dot product in Eq. (\ref{f}) is the symmetric bilinear form 
defined in Eqs. (\ref{4.57}), (\ref{4.58}), (\ref{new})
above for the three cases $h$, $B$ and $\vphi$. Introducing 
Fourier transforms, with the conventions,
\beq
J(p) = \int d^4 x \, e^{-ipx} \, J(x) \, ,
\eeq
\beq
G_{\rm reac} (x) = \frac{1}{2} \, \left [G_{\rm ret} (x) - G_{\rm adv} (x)
\right ] = \int 
\frac{d^4 p}{(2\pi)^4} \, G_{\rm reac} (p) \, e^{+ipx}\,,
\eeq
the energy loss (\ref{f}) reads
\beq
\label{4.79}
P_{\mu}^{\rm lost} = -\,k \int \frac{d^4 p}{(2\pi)^4} \, ip_{\mu} \, G_{\rm reac} 
(p) \, J(-p) \cdot J(p) \, .
\eeq
To see the positivity properties of $P_{\mu}^{\rm lost}$ we need to insert the 
explicit expression of the Fourier transform of $G_{\rm reac}$.

The Fourier decomposition of the retarded and advanced Green functions $(\Box 
\, G = -\delta^4)$ read
\beq
G_{{\rm ret} \atop {\rm adv}} (x) = \int \frac{d^4 p}{(2\pi)^4} \, 
\frac{e^{ipx}}{{\bf p}^2 - (p^0 \pm i \eta)^2} = \int \frac{d^4 p}{(2\pi)^4} \, 
\frac{e^{ipx}}{p^2 \mp i \, \eta \, p^0}\,,
\eeq
where $\eta$ is any positive infinitesimal. Using the formula
\beq
\frac{1}{x-a \pm i \eta} = P \, \frac{1}{x-a} \mp i \pi \delta (x-a)\,,
\eeq
where $P$ denotes the principal part, one finds
\beq
\label{4.82}
G_{\rm reac} (p) = \frac{1}{2} \, [G_{\rm ret} (p) - G_{\rm adv} (p)] = i \pi \ 
\hbox{sign} \, (p^0) \, \delta (p^2) \, .
\eeq
Inserting (\ref{4.82}) into (\ref{4.79}) one gets
\bea
\label{rad}
P_{\mu}^{\rm lost} &=& +\, k\pi \int \frac{d^4 p}{(2\pi)^4} \ \hbox{sign} \, 
(p^0) \, p_{\mu} \, \delta (p^2) \, J(-p) \cdot J(p) \nonumber \\
&=& +\,k \int_{V_+} \widetilde{dp} \, p_{\mu} \, J^* (p) \cdot J(p) \, , 
\eea
where $V_+$ denotes the positive mass shell $p^0 = + \sqrt{{\bf p}^2}$ and 
$\widetilde{dp} = (2\pi)^{-3} \, d^3 {\bf p} / 2p^0$ the natural integration 
measure on $V_+$. Here, we have used the reality of the source: $J^* (x) = J(x) 
\Rightarrow J^* (p) = J(-p)$.

As in the case of Eq. (\ref{momd}) and its kin, the meaning of Eq. (\ref{rad})
is formal when evaluated on a (periodic) free string trajectory. 
However, it is, as usual, 
easy to convert Eq. (\ref{rad}) 
in a result for the average rate of 4-momentum loss by using Fermi's golden rule:
\beq
[\delta (p^0 - n \omega)]^2 = \frac{1}{2\pi} \, \delta (p^0 - n \omega) \int d\tau 
\, .
\eeq
One then recovers known results for the average energy radiation from periodic 
string motions \cite{VilenkinShellard94}, \cite{DamourVilenkin97}.

The integrand in the last result has the good sign (i.e. defines a vector 
within the future directed light cone) if the dot product $J^* (p) \cdot J(p) > 
0$. This is clearly the case for a scalar source, but for the gauge fields 
$h_{\mu \nu}$ and $B_{\mu \nu}$ one has integrands
\beq
\label{ih}
J_{h \mu \nu}^* (p) \, J_h^{\mu \nu} (p) - \frac{1}{2} \ \vert 
J_{h\lambda}^{\lambda} (p) \vert^2 \, ,
\eeq
and
\beq
\label{ib}
J_{B \mu \nu}^* (p) \, J_B^{\mu \nu} (p)
\eeq
which are not explicitly positive because of the wrong sign of the mixed 
components $J_{0i}$. As is well known this potential problem is cured by one 
consequence of gauge invariance, namely some conservation conditions which must 
be satisfied by the source. In our case the gravitational source $J_{\mu \nu}^h 
(x) \propto \widetilde{T}_{\mu \nu} (x)$ must satisfy $\partial^{\nu} \, 
\widetilde{J}_{\mu \nu}^h = 0$, while the axionic source must satisfy 
$\partial^{\nu} \, J_{\mu \nu}^B (x) = 0$. In the Fourier domain this gives 
$p^{\nu} \, \widetilde{J}_{\mu \nu}^h (p) = 0$ or $p^{\nu} \, J_{\mu \nu}^B (p) 
= 0$. These transversality constraints are just enough to ensure that the 
integrands (\ref{ih}), (\ref{ib}) are positive {\it when evaluated on the mass shell} 
$V_+$. What happens in the ``local back reaction approximation'' is that one 
replaces the Green function $G_{\rm reac} (x)$ by a distributional kernel 
$G_{\rm loc} (x)$ with support (in $x$ space) localized at $x=0$. Its Fourier 
transform $G_{\rm loc} (p)$ is no longer localized on the light cone $p^2 = 0$, 
and therefore the delicate compensations ensuring the positivity of the 
integrands (\ref{ih}), (\ref{ib}) do not work anymore.
This explains why the local back reaction approximation is prone 
to giving unreliable expressions for the damping due to {\it gauge} fields. 
On the other hand, in the case of a scalar field the crucial source integrand 
$J^*(p)\,J(p)$ in Eq.~(\ref{rad}) remains positive-definite 
even off the correct mass shell. This explains why, in the case 
of the dilatonic field, the local back reaction 
approximation might (as it was found above to do) define a physically
acceptable approximation to the exact, non-local damping effects.

\section{Improved dilatonic reaction as substitute to gravitational reaction}
\label{sec5}
As the main motivation of the present study is to find a physically reasonable, 
and numerically acceptable, approximation to gravitational radiation damping, 
the results of the previous Section would seem to suggest that the local back 
reaction approach fails to provide such an approximation. However, we wish to 
propose a more positive interpretation. Indeed, both the direct verification of 
Section \ref{subsec4.3}, and the argument (in Fourier space) of 
Section \ref{subsec4.5} shows that 
the local back reaction approximation can make sense when applied to scalar 
fields. On the other hand, Damour and Vilenkin \cite{DamourVilenkin97} 
in a recent study of 
dilaton emission by cosmic strings have found that, in spite of their genuine 
physical differences, gravitational radiation and dilatonic radiation from 
strings are globally rather similar. For the samples of cuspy or kinky loops 
explored in Ref. \cite{DamourVilenkin97}, the global energy losses 
into these fields turned 
out to be roughly proportional to each other. Even when considering in more 
detail the physically important problem of the amount of radiation from cusps, 
it was found that (despite an expected difference linked to the spin 2 
transversality projection) both radiations were again roughly similar.

Let us also recall that this similarity, or better brotherhood, between 
gravitational and dilatonic couplings is technically apparent in the similarity 
of their vertex operators (which are both subsumed in the form $\zeta_{\mu \nu} 
\, \partial^a \, z^{\mu} \, \partial_a \, z^{\nu}$ with a generic symmetric 
polarization tensor $\zeta_{\mu \nu}$) and is a very important element of 
superstring theory. This leads us to propose to use, after a suitable 
normalization, the physically acceptable local dilatonic back reaction force as 
a {\it substitute} for the gravitational radiation one. In other words, we 
propose to use as ``approximation'' to gravitational radiation damping a local 
reaction force of the form (in conformal gauge)
\beq
\label{for}
{\cal F}_{\mu} = \frac{4}{3}G\,\mu^2\,\Delta\,\left [ 
\tdot{z}_\mu - \dot{z}_\mu\,\left ( \frac{\dot{z}\cdot \tdot{z}}{\dot{z}^2} \right ) 
+ {z}^{\prime}_{\mu}\,\left ( \frac{{z}^\prime 
\cdot \tdot{z}}{\dot{z}^2} \right )\right ]\,.
\eeq
We note also that, though there are more differences between axionic and 
gravitational radiations than between the dilatonic and gravitational ones, they 
are still roughly similar in many ways (as witnessed again by the brotherhood 
of their vertex operators $\zeta_{\mu \nu} \, \partial^a \, z^{\mu} \, 
\partial_a \, z^{\nu}$ with now a generic asymmetric polarization tensor) so 
that one can hope to be able also to represent in an acceptable manner axionic 
radiation damping by a force of the type (\ref{for}) with the replacement $G\mu^2 
\rightarrow G\lambda^2$ and another, suitable choice of $\Delta$. [Actually, 
due to their sign error, this last proposal agrees with the practical 
proposition made in Refs.~\cite{BS95}, \cite{BS96}.]

It remains to clarify the choice of $\Delta$ in Eq. (\ref{for}). 
Up to now we have implicitly 
assumed that $\Delta$ was constant. There are, however, several reasons for 
suggesting a non-constant $\Delta$. The first reason concerns energy-momentum 
losses associated with cusps. To see things better, let us use a temporal gauge 
$t=\tau$ and concentrate on the energy loss implied by Eq. (\ref{for}). One finds 
simply
\beq
\label{eloss}
E^{\rm lost} = -\frac{4}{3} \, G \mu^2 \int \int d\sigma \, d\tau \, \Delta \, 
\frac{{\bf v} \cdot \ddot{\bf v}}{1- {\bf v}^2} \, ,
\eeq
where ${\bf v} (\sigma , \tau) \equiv \dot{\bf z} (\sigma , \tau)$. At a cusp 
${\bf v}^2 (\sigma , \tau) = 1$. As ${\bf v}^2 (\sigma , \tau) \leq 1$ 
everywhere, near a cusp one will have ${\bf v}^2 (\sigma , \tau) = 1 - 
(a\sigma^2 + b\sigma \tau + c\tau^2) + {\cal O} ((\sigma + \tau)^3)$ where the 
parenthesis is a positive definite quadratic form. This shows that, if $\Delta$ 
is constant, the integral $E^{\rm lost} \sim \int \int d\sigma \, d\tau \, 
(1-{\bf v}^2)^{-1}$ is logarithmically divergent (as we explicitly verified on 
specific string solutions). As the real energy loss to gravitational or 
dilatonic radiation from (momentary) cusps is finite, this shows that Eq. (\ref{for}) 
overestimates the importance of back reaction due to cusps. In other words, if 
one tries to complete the equations of motion of a string by adding the force 
(\ref{for}) with $\Delta =$ const, this reaction force will prevent the appearance of 
real cusps. As the calculations of Ref. \cite{QuashnockSpergel90}, using the 
``exact'' non-local gravitational radiation, find that cusps are weakened but 
survive, it is clear that one must somehow soften the ``local'' force (\ref{for}) if 
we wish to represent adequately the physics of cusps. At this point it is 
important to note that the proposal (\ref{for}) lacks worldsheet covariance, 
which means, on the one hand, that $\Delta$ has introduced a local coordinate 
length or time scale on the 
worldsheet, rather than an invariant interval and, on the other hand, that one 
must specify a particular time-slicing of the worldsheet. 
As the ratio between coordinate 
lengths and times and proper intervals is locally given by the square root of 
the conformal factor $e^{\phi} = z'^2 = - \dot{z}^2$ ($= 1 - {\bf v}^2$ in 
temporal gauge), it is natural to think that a better measure of the coordinate 
interval $\Delta$ to use in Eq. (\ref{for}) might vary along the worldsheet because 
it incorporates some power of $e^{\phi}$. This might (if this power is 
positive) prevent the logarithmic divergence of the integral (\ref{eloss}). At this 
stage, a purely phenomenological proposal is to take $\Delta$ in Eq. (\ref{for}) of 
the form
\beq
\label{del}
\Delta (\sigma , \tau) = f \, (- \dot{z}^2)^{\eta} \, 2 \lambda\,,
\eeq
where $f$ is a dimensionless factor, $\eta$ is a positive power, and $\lambda$ 
the wavelength of the radiatively dominant mode emitted by the string. We 
introduced a factor two for convenience because, in the case of loops for which 
the fundamental mode is dominant, the wavelength is $L/2$ where $L$ is the 
invariant length of the loop. On the other hand, if we consider a loop carrying mainly 
high-frequency excitations, or an infinite string, it is clear that $\Delta$ 
should not be related to the total length $L$, but to a length linked to the 
scale of the principal modes propagating on the string.

Let us briefly comment on the lack of worldsheet covariance of Eq.~(\ref{for}) 
and on its consequences.  Eq.~(\ref{for}) emerged as a local approximation to an 
integral which had the same formal expression in all conformal gauges. 
The ``local back reaction approximation'' procedure 
has, among other things, violated the formal symmetry between $\tau$ and $\sigma$ 
on the worldsheet. From the formal point of view this loss of symmetry 
is certainly unpleasant and it would be nicer to be able to write a local force 
density which respects the symmetry of the worldsheet conformal gauges, and 
does as well as Eq.~(\ref{for}) in entailing a positive energy loss quantitatively 
comparable to the result (\ref{eloss}) (which will be seen below to be an adequate 
representation of the actual energy loss). We failed to find such a covariant 
local force density. This is why we propose to use (\ref{for}), despite its formal 
imperfections, as a substitute to the exact, non-local gravitational 
radiation damping. From this point of view, 
the asymmetry between $\tau$ and $\sigma$ in Eq.~(\ref{for}) can be interpreted 
as a sign that the purely local expression (\ref{for}) tries its best to incorporate 
the, in reality, global damping effects 
by selecting special time-slicings of the worldsheet ($\tau = {\rm const.}$ 
lines, and their orthogonal trajectories). A natural physical choice of special time-slices 
(necessary to define properly the meaning of (\ref{for})) is to consider the spatial 
sections associated to the (instantaneous) center of mass-frame of the string. 
[Note that the numerical calculations below of the energy loss (\ref{eloss})
are performed in the string center of mass frame.]
For a free Nambu-Goto string in flat space, this definition is compatible 
with using a worldsheet gauge which is both conformal and temporal 
(i.e. $\tau \propto P_\mu^{\rm string}\,z^\mu$). 
Therefore, in such a case, the (orthogonal) worldsheet vector field 
$\pa/\pa \tau$ is well defined (both in direction and in normalization), 
which means that ${\cal F}_\mu$, Eq.~(\ref{for}) is well defined, 
on the worldsheet, as a spacetime vector locally orthogonal 
to the worldsheet. We shall admit that the definition 
of ${\cal F}_\mu$ can be smoothly extended to the case where 
the (Nambu) string moves in a curved background spacetime 
(say a Friedmann universe). When working in the approximation 
of a flat background the expression (\ref{for}) can be used directly 
as right-hand side of the standard, flat-space, conformal gauge string equations 
of motion: $\eta_{\mu \nu}\, (\ddot{z}^\nu - z^{\prime \prime \, \nu}) = 
{\cal F}_\mu$. [Note that we use here a flat-space worldsheet gauge, 
Eq.~(\ref{con}) with 
$g_{\mu \nu} \rightarrow \eta_{\mu \nu}$.] Note finally that the actual numerical 
simulations of a string network introduce a particular time slicing 
and one might also decide (for pragmatic reasons) to use it to define the 
$\tau = {\rm const.}$ slices of Eq.~(\ref{for}) (i.e. to neglect 
the Lorentz-transformation effects associated to the center of mass-motion 
of the strings).

A first check of the physical consistency of the proposal (\ref{del}) consists in 
verifying that, despite the nonconstancy of $\Delta$, the integrated energy 
loss (\ref{eloss}) will be positive for all possible loop trajectories. Integrating 
by parts Eq. (\ref{eloss}) one finds
\beq
E^{\rm lost} = \frac{4}{3} \, G\,\mu^2\,f (2\lambda) \int \int d\sigma \, d\tau \, 
(1-{\bf v}^2)^{\eta} \, \left[ \frac{\dot{\bf v} \cdot \dot{\bf v}}{1-{\bf 
v}^2} + 2(1-\eta) \, \frac{({\bf v} \cdot \dot{\bf v})^2}{(1-{\bf v}^2)^2} 
\right] \, .
\eeq
This is manifestly positive (and finite) as long as $0 < \eta < 1$. Assuming 
this to be the case, the question is then: Are there values of $f$ and $\eta$ 
(after having decided on a precise definition of $\lambda$) such that the 
corresponding damping force (\ref{for}) gives a reasonably accurate description of 
the ``exact'' effects of energy loss to gravitational radiation? We did not try 
to answer this question in full generality. For simplicity, we fixed the power 
$\eta$ to the value $\eta = \frac{1}{2}$ (which seems intuitively preferred as 
it evokes a Lorentz contraction factor arising because we look at string 
elements ``moving'' with relativistic speeds). Then we compared the energy loss 
due to (\ref{for}) to the energy radiated in gravitational waves as (computed (using 
Eq. (\ref{rad})) in the literature (both energy losses being evaluated 
in the rest frame of a free string). 
As a sample of loop trajectories we consider 
Burden loops \cite{Burden85}
\beq
{\bf z} (\tau , \sigma) = \frac{1}{2} \, [{\bf a}(u) + {\bf b}(v)]\,, 
\eeq
\bea
{\bf a} &=& \frac{L}{2 \pi}\, \left [\frac{1}{m}\cos (m\,u)\,\vec{e}_3 + 
\frac{1}{m}\sin (m\,u)\,\vec{e}_1 \right ]\,,\\
{\bf b} &=& \frac{L}{2 \pi}\, \left [\frac{1}{n}\cos (n\,v)\,\vec{e}_3 - 
\frac{1}{n}\sin (n\,v)\,\vec{e}_1^{\,\prime} \right ]\,,
\eea
where 
\beq
u = \frac{2 \pi}{L}\,(\tau-\sigma)\,, \quad \quad   
v = \frac{2 \pi}{L}\,(\tau+\sigma)\,, \quad \quad 
\vec{e}_1^{\,\prime} = \cos \psi\, \vec{e}_1 + \sin \psi\, \vec{e}_2\,. 
\eeq
This family of solutions depends on the overall scale $L$, which is the total 
invariant length of the loop $(M=\mu L)$, on two integers $m$ and $n$, and on 
the angle $\psi$. Our parameter $\psi$ coincides with the angle $\psi$ in 
\cite{Burden85}, denoted $\varphi$ in \cite{VilenkinShellard94}. The actual oscillation 
period of the loop is $T = L/(2mn)$ which leads us to choosing $2\lambda = 2T = 
L/mn$ in Eq. (\ref{del}). With this choice we computed the energy loss (\ref{eloss}). The 
calculation is simplified by noting, on the one hand, that, for this family of 
loops, ${\bf v} \cdot \ddot{\bf v} = - \left (\frac{2 \pi}{L}\right )^2\,
\left( \frac{m^2 + n^2}{2} \right) \, 
{\bf v}^2$, and on the other hand that the worldsheet integral in (\ref{eloss}) can 
be rewritten in terms of an average over linear combinations of 
the two angles $2\pi m (\tau - \sigma) 
/ L$ and $2\pi n (\tau + \sigma) / L$. This yields simply for the average rate 
of energy loss
\beq
\Gamma_{m,n} \equiv \frac{\dot{E}^{\rm lost}}{G\mu^2} = \frac{4}{3} \, f \, 
\frac{m^2 + n^2}{2mn} \, \gamma \, ,
\eeq
where
\beq
\gamma = 4\int_0^{\pi}d x \,\int_0^\pi dy 
\left [ -\sqrt{1-{\bf v}^2} + \frac{1}{\sqrt{1-{\bf v}^2}} \right ]\,,
\eeq
with
\beq
{\bf v}^2 = \frac{1}{2}\,\left [ 1 - \frac{1}{2}\,( 1+\cos \psi)
\,\cos x -\frac{1}{2}\,( 1-\cos \psi)
\,\cos y \right ]\,.
\eeq
\begin{figure}
\centerline{\epsfig{file=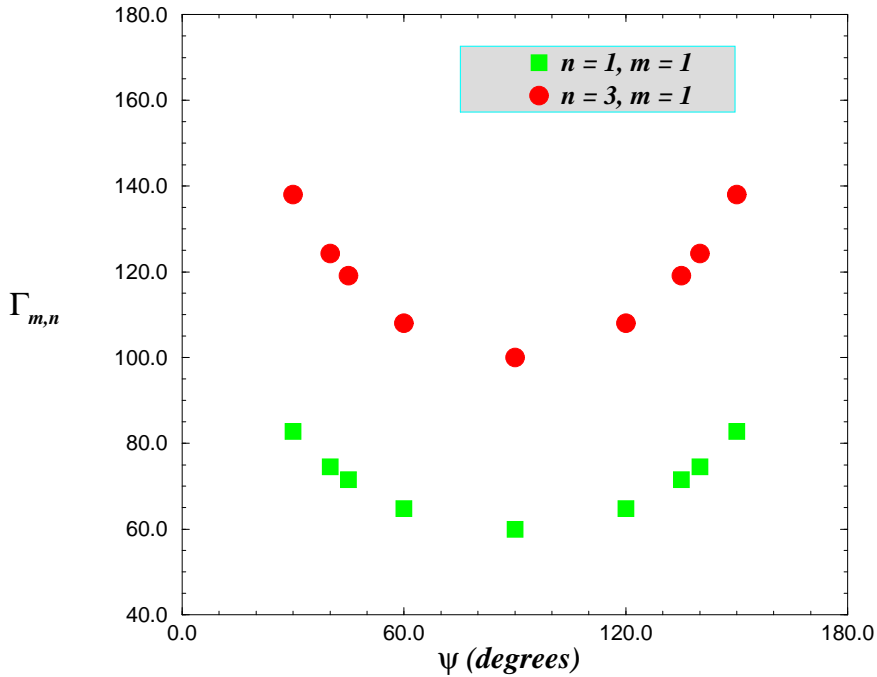,width=0.65\textwidth,angle=-90}}
\caption{\sl Dimensionless energy loss rate Burden loops with 
$(m,n) =(1,1)$ and $(1,3)$.}
\label{fig1}
\end{figure}
We plot in Fig. \ref{fig1} $\Gamma_{m,n}$ as a function of the angle $\psi$, for the 
nominal value $f=1$ and for the two cases $(m,n) = (1,1)$, $(m,n) = (1,3)$. 
[As said above there is a simple scaling law for the dependence on $m$ and 
$n$.] If one compares this Figure with the figures published in 
\cite{Burden85}, \cite{VilenkinShellard94} (Fig. 7.6, p. 205 there)
one sees that they give a roughly adequate 
numerical representation of energy losses to gravitational radiation if
\beq
f \simeq 0.8 \, .
\eeq
The fact that our present ``best fit'' value of the factor $f$ leads to values 
of $\Delta$ which are numerically comparable to $L$ (when $(m,n) = (1,1)$) 
rather than to a smaller fraction of $L$ should not be considered as physically 
incompatible with the idea of using a local approximation to back reaction. 
Indeed, on the other hand, the rough justification of the local approximation 
given in Section \ref{sec4} suggested $\Delta \sim \pi / \omega \sim \lambda / 2$, i.e. 
something like $L/4$, and, on the other hand, numerical computations show that 
the energy lost to dilaton waves (with coupling $\alpha = 1$) is {\it smaller} 
than that lost to gravitational waves by a factor of order 3 or so (part of 
which is simply due to the fact that there are two independent tensor modes 
against one scalar mode). Therefore, as we use $\Delta$ only as an {\it 
effective} parameter to model gravitational damping it is normal to end up with 
an increased value of $\Delta / L$.

Clearly, more work would be needed to confirm that the modified local dilaton 
reaction (\ref{for}) can be used as a phenomenological representation of 
gravitational reaction. Our main purpose here was to clarify the crucial sign 
problems associated to gauge fields, and to give a first bit of evidence 
indicating that Eq. (\ref{for}) deserves seriously to be considered as an interesting 
candidate for mimicking, in a computationally non-intensive way, the back 
reaction of gravitational radiation. We are aware that several important issues 
will need to be further studied before being able to use Eq. (\ref{for}) in a network 
simulation. Some numerically adequate definition of $\lambda$ will have to be
provided beyond a case by case definition, which in the case of long loops 
decorated by a regular array of kinks, as in Ref. \cite{QuashnockSpergel90}, 
would be something like $2\lambda \sim L/N$ where $N$ is the total number of 
kinks. We note in this respect that a Burden loop with $m=1$ and 
$n \gg 1$ provides a simple model of a long, circular loop decorated 
by a travelling pattern of small transverse oscillations. 
However, the local approximation (\ref{for}) cannot expected 
to be accurate in this case, because the radiation from purely 
left-moving or right-moving modes is known to be suppressed 
\cite{VilenkinShellard94}. This suppression is not expected to hold 
in the more physical generic case where the transverse oscillations move both ways.
The accuracy of the local approximation (\ref{for}) should 
therefore be tested only in such more generic cases. 

The explicit expression (\ref{for}) must be rewritten in the temporal, but 
not necessarily conformal, worldsheet gauges used in numerical simulations, and 
the higher time derivatives in ${\cal F}_{\mu}$ must be eliminated by using (as 
is standard in electrodynamics \cite{LandauLifshitz} and gravitodynamics 
\cite{Damour83}) the lowest-order equations of motion. [These last two issues 
have already been treated in Refs. \cite{BS95}, \cite{BS96}.] Finally, we did 
not try to explore whether $\eta = 1/2$ is the phenomenologically preferred 
value. To study this point one should carefully compare the effects of (\ref{for}) on 
the weakening of cusps and kinks with the results based on the exact, non-local 
reaction force \cite{QuashnockSpergel90}. [The facts that the curves 
in Fig. \ref{fig1} are flatter than the corresponding figures 
in \cite{Burden85}, \cite{VilenkinShellard94} suggest that a smaller value of $\eta$ might give a better fit.]

\section{Conclusions}
\label{sec6}
In this paper we studied the problem of the radiation reaction on cosmic strings caused 
by the emission of gravitational, dilatonic and axionic fields. 
We assume the absence of external fields. We use a straightforward 
perturbative approach and work only to first order in $G$.
Our main results are the following.
\begin{itemize} 
\item Using the results of Refs.~\cite{CB}, \cite{BDletter} for the renormalization 
of the string tension $\mu$, we write down the explicit form, at linear order in $G$, 
of the renormalized equations of motion
of a string interacting with its own 
(linearized) gravitational, dilatonic and axionic 
fields. [Within our framework, we verified the on shell finiteness of the bare 
equations of motion, which is equivalent to their renormalizability.]

\item We have extended a well-known result of Dirac by proving 
for general linearized fields that, 
in the decomposition (\ref{4.9}) of the renormalized self-force, 
only the time-antisymmetric contribution ${\cal F}_\mu^{\rm reac} = 
{\cal F}_\mu(A_{\rm reac})$, where $A_{\rm reac}(x)$ is 
the half-retarded minus half-advanced field, contributes, 
after integration over time, to the overall damping of the source.
[This result had been assumed without proof in previous 
work on the topic.] The ``reactive'' self-force ${\cal F}_\mu^{\rm reac}$ 
is manifestly finite (and independent of the renormalization length 
scale $\Delta_R$), and is non-local.

\item We have critically examined the proposal of Battye and Shellard 
\cite{BS95}, \cite{BS96} (based on an analogy with the Abraham-Lorentz-Dirac 
treatment of self-interacting point charges) to approximate the non-local 
integral (\ref{4.10}) entering the reactive self-force ${\cal F}_\mu^{\rm reac}$ 
by the local expression (\ref{locapp}). For this purpose we found very convenient 
to use {\it dimensional continuation}, a well known technique in quantum field 
theory. We found that the local back reaction approximation gives 
antidamping for the axionic field, and a vanishing net 
energy-momentum loss for the gravitational one. We argued that the ultimate 
origin of these physically unacceptable results come from trying 
to apply the local back reaction approximation to {\it gauge } fields. 
The non-positivity of the local approximation 
to the damping comes from combining the modification 
of the field Green functions implicit in the local back reaction 
method, with the delicate sign compensations ensured, 
on shell only, by the transversality constraints of the sources 
of gauge fields.
\item By contrast, we find that the local approximation 
to the dilatonic reaction force has 
the correct sign for describing a radiation 
damping. In the case of a non-gauge 
field such as the scalar dilaton there are no delicate sign 
compensations taking place, and  
the coarse approximation of the field Green function,  
implicit in the local back reaction method, can (and does) lead to physically 
acceptable results. 
\item  Taking into account the known similarity between the gravitational and dilatonic
radiations (e.g. \cite{DamourVilenkin97}), we propose to use as effective substitute 
to the exact (non-local) gravitational radiation damping 
the ``dilaton-like'' local reaction force (\ref{for}), with a suitably 
``redshifted'' effective length $\Delta$, Eq. (\ref{del}). This force 
is to be used in the right-hand side of the standard, flat-space 
conformal-world-sheet-gauge string equations of motion, with $\tau$-
slicing linked, say, to the global center-of-mass frame of the string. 
The numerical calculations exhibited in Fig.~\ref{fig1} give some evidence 
indicating that Eq.~(\ref{for}) deserves seriously to be considered 
as an interesting candidate for {\it phenomenologically approximating}, 
in a computationally non-intensive way, the back reaction of gravitational 
radiation. [We recall that the exact, non-local approach to gravitational 
back reaction, defined by Eq.~(\ref{4.9}), is numerically so demanding that there is
little prospect to implementing 
it in full string-network calculations.] More work is needed (e.g. by 
comparing the dynamical evolution of a representative sample of 
cosmic string loops under the exact renormalized self-force (\ref{4.9}) 
and our proposed (\ref{for})) to confirm that our proposed substitute 
(\ref{for}) is a phenomenologically acceptable representation of gravitational 
reaction (or of the combined dilatonic-gravitational reaction, 
as string theory suggests that the dilaton is a model-independent 
partner of the Einstein graviton). 
\end{itemize}
It will be interesting to see what are the consequences 
of considering the effective reaction force, Eq.~(\ref{for}), 
in full-scale network simulations (done for several different values of 
$G\mu$) of gravitational radiation.
 Until such simulations (keeping track of the damping of small scale 
structure on long strings) are performed, one will not be able to give any 
precise prediction for the amount and spectrum of stochastic gravitational 
waves that the forthcoming LIGO/VIRGO network of interferometric detectors,
possibly completed by cryogenic bar detectors,
might observe.

\section*{Acknowledgements}
We are grateful to Bruce Allen, Richard Battye, Francois Bouchet, 
Brandon Carter and Alex Vilenkin 
for useful exchanges of ideas. 

\appendix
\section{}
\label{app}
In this Appendix we will give some details on the derivation of 
Eqs.~(\ref{4.32}), (\ref{4.33}) using dimensional continuation. 

A nice feature of analytic continuation is that it allows one to work ``as 
if'' many singular terms were regular. For instance, the factors 
$(\Omega^2)^{(2-n)/2}$ and $(\Omega^2)^{-n/2}$
that appear in Eqs.~(\ref{ret1}), (\ref{dret1}) 
blow up on the light cone 
$(\Omega^2 = 0)$ when $n=4$. However, if we take the real part of $\epsilon = 
4-n$ large enough (even so large as corresponding to negative values for ${\rm 
Re} (n)$), these $\Omega$-dependent factors become finite, and actually {\it 
vanishing}, on the light cone. This remark allows one to deal efficiently with 
the $\Omega$-dependent factors appearing in Eqs.~(\ref{ret1}), 
(\ref{dret1}). We are here interested in 
the contributions to $A_{\rm ret} (z)$ and $\partial_{\mu} \, A_{\rm ret} (z)$ 
coming from a small neighbourhood $z' = z(\tau' , \sigma')$ of $z=z(\tau , 
\sigma)$ on the worldsheet. Let us, for simplicity, denote $\omega \equiv 
\Omega^2$. We first remark that when $(\tau' , \sigma') \rightarrow (\tau , 
\sigma)$, $\omega = - (z(\tau , \sigma) - z(\tau' , \sigma'))^2$ admits an 
expansion in powers of $\tau' - \tau$ and $\sigma' - \sigma$ of the form
\beq
\label{w}
\omega = \omega_2 + \omega_3 + \omega_4 + \ldots \, ,
\eeq
with $\omega_2 = -{\dot z}^2 \, [(\tau' - \tau)^2 - (\sigma' - \sigma)^2]$, 
and
\beq
\omega_3 = {\cal O} \, ((\tau' - \tau)^3 + (\tau' - \tau)^2 \, (\sigma' - 
\sigma) + (\tau' - \tau) \, (\sigma' - \sigma)^2 + (\sigma' - \sigma)^3) \, , 
\ \hbox{etc} \ldots
\eeq
Then we can formally expand the $\Omega$-dependent factors of Eqs.~(\ref{ret1}), 
(\ref{dret1})  in powers 
of $\tau' - \tau$ and $\sigma' - \sigma$ as follows
\bea
T \, [\omega^{\alpha} \, \theta (\omega)] &=& \left[ \omega_2^{\alpha} + \alpha 
\, \omega_2^{\alpha - 1} (\omega_3 + \omega_4 + \ldots) + \frac{\alpha (\alpha 
- 1)}{2} \, \omega_2^{\alpha - 2} (\omega_3 + \ldots)^2 + \ldots \right] \times
\nonumber \\
&& \biggl[ \theta (\omega_2) + \delta (\omega_2) \, (\omega_3 + \omega_4 + 
\ldots) + \delta' (\omega_2) \, (\omega_3 + \ldots)^2 + \ldots \biggl] \, . 
\nonumber
\eea
Here and below, the symbol $T$ will be used to denote a (formal) Taylor 
expansion of any quantity following it. This expansion is valid (at any finite 
order) when ${\rm Re} (\alpha)$ is large enough, and is therefore valid (by 
analytic continuation) in our case where $\alpha = (2-n)/2$ or $-n/2$. A 
technically very useful aspect of the above expansion is that all the terms 
containing $\delta (\omega_2)$ or its derivatives give {\it vanishing} 
contributions (because $\omega_2^{\alpha - k} \, \delta^{(\ell)} (\omega_2)$ 
vanishes if ${\rm Re} (\alpha)$ is large enough, so that, by analytic 
continuation, $\omega_2^{\alpha - k} \, \delta^{(\ell)} (\omega_2) = 0$ for all 
values of $\alpha$). The net effect is that the contribution coming from a 
small string segment $-\frac{\Delta}{2} < (\sigma' - \sigma) 
< \frac{\Delta}{2}$ 
around $\sigma$ (with $\Delta$ being much smaller that the local radius of 
curvature of the worldsheet) can be simply (and correctly) written as the 
following expansion:
\bea
[A_{\rm ret} (z)]^{\Delta} &\equiv& \epsilon \int_{\sigma -\Delta / 2}^{\sigma+\Delta / 2} 
d\sigma' \int_{-\infty}^{\tau_1} d\tau' \, \Sigma \, \omega^{\frac{2-n}{2}} \, 
\theta (\omega) \nonumber \\
&=& \epsilon \int_{\sigma-\Delta / 2}^{\sigma+\Delta / 2} d\sigma' \int_{-\infty}^{\tau_1} 
d\tau' \, T \, \left(\Sigma \, \omega^{\frac{2-n}{2}}\right) \, \theta 
(\omega_2) \nonumber \\
&=& \epsilon \int_{\sigma-\Delta / 2}^{\sigma+\Delta / 2} d\sigma' \int_{-\infty}^{\tau - 
\vert \Delta \sigma \vert} d\tau' \, T \, \left(\Sigma \, 
\omega^{\frac{2-n}{2}}\right) \, . \nonumber
\eea
Here, we have introduced an arbitrary upper limit $\tau_1$, submitted only to 
the constraint $\tau_{\rm ret} < \tau_1 < \tau_{\rm adv}$ (for instance 
$\tau_1$ could be $\tau$), and which replaces the missing theta function $\theta 
(z^0 - z'^0)$ by selecting the retarded portion of the other theta function 
$\theta (\omega)$. As above, the symbol $T$ denotes a formal Taylor expansion. 
The expansion $T (\Sigma \,\omega^{\alpha})$ is simply obtained by multiplying 
the expansion (\ref{w}) of $\omega$ with that of $\Sigma (\tau' , \sigma')$, namely
\beq
T \, [\Sigma (\tau' , \sigma')] = \Sigma (\tau , \sigma) + (\tau' - \tau) \, 
\dot \Sigma + (\sigma' - \sigma) \, \Sigma' + \ldots
\eeq
Similarly we have
\beq
[\partial_{\mu} \, A_{\rm ret} (z)]^{\Delta} = 2 \epsilon \int_{\sigma-\Delta / 
2}^{\sigma+ \Delta / 2} d\sigma' \int_{-\infty}^{\tau - \vert \Delta \sigma \vert} 
d\tau' \, T \, (\Sigma \, \Omega_{\mu} \, \omega^{-n/2}) \, ,
\eeq
as well as corresponding expressions for the advanced fields
\bea
[A_{\rm adv} (z)]^{\Delta} &=& \epsilon \int_{\sigma-\Delta / 2}^{\sigma+\Delta / 2} d\sigma' 
\int_{\tau + \vert \Delta \sigma \vert}^{+\infty} d\tau' \, T \, 
\left(\Sigma \, \omega^{\frac{2-n}{2}}\right ) \,,
\eea
\bea
[\pa_\mu \, A_{\rm adv} (z)]^{\Delta} &=& 2 \epsilon 
\int_{\sigma-\Delta / 2}^{\sigma+\Delta / 2} d\sigma' \,
\int_{\tau + \vert \Delta \sigma \vert}^{+ \infty} d\tau'\, 
T \, \left (\Sigma \, \Omega_{\mu} \, \omega^{-n/2} \right ) \,.
\eea
As a check, we first computed the ultraviolet divergent 
contributions to $A_{\rm ret} (z)$ and $\partial_{\mu} \, A_{\rm ret} (z)$. We 
find 
\bea
\label{fiedim}
A_{\rm ret}(z) &=& -\frac{1}{\dot{z}^2}\,\left (\frac{2}{\epsilon}\right)\,2\,\Sigma\,,\\
\label{derfiedim}
\pa_\mu A_{\rm ret}(z) &=& 
\frac{1}{(\dot{z}^2)^2}\,\left (\frac{2}{\epsilon}\right )\,
\left [ - \Sigma\,\ddot{z}_\mu + \Sigma\,z_{\mu}^{\prime \prime} + 4\Sigma\,z_\mu^\prime\,
\left ( \frac{z^{\prime}\cdot z^{\prime \prime}}{\dot{z}^2}\right)
+4\Sigma\,\dot{z}_\mu\,\left (\frac{\dot{z}\cdot \ddot{z}}{\dot{z}^2}\right ) \right .
\nonumber \\
&& + \left . 2 \Sigma^\prime\,z^\prime_\mu - 2\dot{\Sigma}\,\dot{z}_\mu \right ]\,.
\eea
As it should, Eq. (\ref{derfiedim}) yields exactly the same divergences as we found 
in Sec. \ref{sec3} 
by introducing a cut-off $\delta$ in the $\sigma'$ integration in four 
dimensions. More precisely, Eq. (\ref{derfiedim}) coincides with Eq.~(\ref{delder}) 
if we change 
$2/\epsilon \rightarrow \log \, 1/\delta$. Let us note that, in the present 
approach, the renormalization scale $\Delta_{\rm R}$ would enter by being introduced 
as a dimension-preserving factor in the dimensionful coupling constants, 
like Newton's constant $G$, say $G^{(n)} = G^{(n=4)} \, \Delta_{\rm R}^{\alpha}$.

Our main interest is to compute the ``local approximations'' to the reaction 
field
\beq
A_{\rm reac}(x) = \frac{1}{2}\,\left ( A_{\rm ret}(x)- A_{\rm adv}(x) \right )\,,
\eeq
and its derivatives. Dimensional continuation gives an efficient tool for 
computing these. Indeed, combining the previous expansions we can write
\bea
\label{rad1}
A_{\rm reac}(x) &=&-\epsilon\,\int_{-\Delta/2}^{\Delta/2} 
d \sigmap \int_{\tau +|\Delta\sigma|}^{+\infty}
d \taup \,\theta(\Omega_0^2)\,T_{(\taup -\tau)\,\mbox{\tiny\rm odd}}
\left(\Sigma\,(\Omega^2)^{\frac{2-n}{2}}\right )\,,\\
\label{drad1}
\pa_\mu A_{\rm reac}(x) &=&
-2\epsilon\,\int_{-\Delta/2}^{\Delta/2} d \sigmap \int_{\tau +|\Delta\sigma|}^{+\infty}\,
d \taup \,\theta(\Omega_0^2) \,T_{(\taup -\tau)\,\mbox{\tiny\rm odd}}
\left ( \Sigma\,\Omega_\mu\,
(\Omega^2)^{-\frac{n}{2}}\right )\,, 
\eea
where $T_{(\tau' - \tau) \, {\rm odd}}$ denotes the part of the Taylor 
expansion which is odd in $\tau' - \tau$. Moreover, as we know in advance (and 
easily check) that the $\sigma'$-integrands in Eqs.~(\ref{rad1}) and 
(\ref{drad1}) are regular at $\sigma' = 
0$, we can very simply write the result of the local approximation
(\ref{locapp}) (with a corresponding definition for $A_{\rm reac}^{\rm local} (z)$) 
by replacing $\sigma' 
= \sigma$ in the integrands of Eqs. (\ref{rad1}), (\ref{drad1})
\beq
\label{y1}
[A_{\rm reac} (z)]^{\rm local} = -\epsilon \Delta \int_{\tau}^{+\infty} d\tau' 
\, T_{(\tau' - \tau) \, {\rm odd}}^{\sigma' = \sigma} \, \left[ \Sigma 
(\Omega^2)^{\frac{2-n}{2}} \right] \, ,
\eeq
\beq
\label{y2}
[\partial_{\mu} \, A_{\rm reac} (z)]^{\rm local} = -2 \epsilon \Delta 
\int_{\tau}^{+\infty} d\tau' \, T_{(\tau' - \tau) \, {\rm odd}}^{\sigma' = 
\sigma} \, \left[ \Sigma \, \Omega_{\mu} \, (\Omega^2)^{-\frac{n}{2}} \right] \,.
\eeq
Here, $T_{(\tau' - \tau) \, {\rm odd}}^{\sigma' = \sigma}$ denotes the 
operation of replacing $\sigma'$ by $\sigma$ and keeping only the odd terms in 
the remaining Taylor expansion in $\tau' - \tau$. This simplifies very much the 
computation of the reactive terms (making it only a slight generalization of 
the well known point-particle results, as given for a general source in, e.g. 
\cite{Damour75}). 
Indeed, inserting the following expansions 
\bea
\Omega_{\mu}(\taup,\sigma) &\simeq& - (\taup-\tau)\,\dot{z}_\mu 
-\frac{1}{2}(\taup-\tau)^2\,\ddot{z}_\mu - \frac{1}{6}(\taup-\tau)^3\,
\tdot{z}_\mu\,, \\
 \Sigma(\taup,\sigma) &\simeq& \Sigma(\tau,\sigma) + (\taup-\tau)\,\dot{\Sigma} + 
+ \frac{1}{2}(\taup-\tau)^2\,\ddot{\Sigma}\,, \\
\Omega^2(\taup,\sigma) &\simeq& -\dot{z}^2\,(\taup-\tau)^2\,\left [
1 + {(\taup-\tau)}\,
\left (\frac{\dot{z} \cdot \ddot{z}}{\dot{z}^2} \right ) + 
\frac{1}{4}\,{(\taup-\tau)^2}\,
\left (\frac{\ddot{z} \cdot \ddot{z}}{\dot{z}^2} \right ) + \right. 
\nonumber \\ 
&& \left . \frac{1}{3}\,{(\taup-\tau)^2}\,
\left (\frac{\tdot{z} \cdot \dot{z}}{\dot{z}^2} \right )
\right ]\,,\vrsmall
\eea
in Eqs.~(\ref{y1}), (\ref{y2}) we get our main results
\bea 
\label{ar}
{[A_{\rm reac}(z)]}^{\rm local} &=& \frac{\Delta}{\dot{z}^2}\,\left [
\dot{\Sigma} - \Sigma\,\left (\frac{\dot{z} \cdot 
\ddot{z}}{\dot{z}^2} \right )\right] \,,\\
\label{dar}
{[\pa_\mu A_{\rm reac}(z)]}^{\rm local} &=& 
\frac{\Delta}{(\dot{z}^2)^2}\,\left [ 
\frac{1}{3}\Sigma\,\tdot{z}_\mu + 
\dot{\Sigma}\,\ddot{z}_\mu + \ddot{\Sigma}\,\dot{z}_\mu - 
4\dot{\Sigma}\,\dot{z}_\mu \, 
\left ( \frac{\dot{z}\cdot \ddot{z}}{\dot{z}^2} \right )
-2 \Sigma \ddot{z}_\mu\, \left ( \frac{\dot{z}\cdot \ddot{z}}{\dot{z}^2} 
\right )\right .\nonumber \\
&& \left . -\Sigma\,\dot{z}_\mu \,\left ( \frac{\ddot{z}\cdot 
\ddot{z}}{\dot{z}^2} \right )
- \frac{4}{3}\,\Sigma\,\dot{z}_\mu\,\left ( \frac{\dot{z} \cdot 
\tdot{z}}{\dot{z}^2} \right )
+ 6 \Sigma\,\dot{z}_\mu\,\left ( \frac{\dot{z}\cdot \ddot{z}}{\dot{z}^2} \right )^2
\right ]\,.
\eea
These results were also obtained (as a check) from Eqs.~(\ref{rad1}), (\ref{drad1})
without using in advance the simplification of putting $\sigmap=\sigma$ 
in the integrand.

We have also performed a direct check on these final expressions by comparing them   
to the well known point-particle case \cite{Fremberg}, \cite{Ma}, \cite{Damour75}.
Indeed, we have seen above that $A_{\rm reac}^{\rm local}$ and 
$\pa_\mu A_{\rm reac}^{\rm local}$ could be thought of as being 
generated by the effective source  
$\Sigma^{\rm eff.}(\taup,\sigmap) = \delta(\sigmap-\sigma)\,\Delta\,\Sigma(\taup,\sigma)$, 
i.e. a source along the world-line ${\cal L}_\sigma$, defined by $\sigmap=\sigma$. 
For any given value of $\sigma$, by transforming  
the coordinate time $\taup$ into the proper time 
$s=\int e^{\phi/2}\,d \taup$ along ${\cal L}_\sigma$
 and by renormalizing 
in a suitable way the source $\Delta\,\Sigma(\taup,\sigma)\equiv e^{\phi/2}\,\tilde{S}(s)$
( so that the stringy spacetime source $\int d^2 \sigmap \Sigma^{\rm eff.}(\taup,\sigmap) 
\,\delta^4(x-z(\sigmap))$ transforms into the standard point-particle source 
$\int ds\, \tilde{S}(s)\,\delta^4(x-z(s))$), we recovered from 
Eqs.~(\ref{ar}), (\ref{dar}) known point-particle results \cite{Damour75}. 
This check is powerful enough to verify the correctness of all the coefficients 
in Eqs.~(\ref{ar}), (\ref{dar}).

In order to compare directly our expressions with what derived by 
Battye and Shellard 
in \cite{BS95}, \cite{BS96}, let us write Eq.~(\ref{dar}) 
for the axion field. We get 
\beq
\label{eqh}
H^{\lambda \mu \nu} = \frac{4G\,\lambda\,\Delta}{(\dot{z}^2)^2}\,
\left [ \frac{1}{3}\,\tdot{z}^{\left [\lambda \right .}\,V^{\left .\mu \nu\right ]}
+ \ddot{z}^{\left [ \lambda \right. }\,\dot{V}^{\left .\mu \nu\right ]} + 
\dot{z}^{\left [\lambda \right .}\,\ddot{V}^{\left .\mu \nu\right ]} -
4\dot{z}^{\left [ \lambda \right. }\,\dot{V}^{\left .\mu \nu \right ]}\,
\left ( \frac{\dot{z}\cdot \ddot{z}}
{\dot{z}^2}\right ) 
- 2 \ddot{z}^{\left [\lambda\right .}\,
{V}^{\left .\mu \nu \right ]}\,\left ( \frac{\dot{z}\cdot \ddot{z}}
{\dot{z}^2}\right )\right ]\,, 
\eeq
where $K^{[\lambda \mu \nu]} = K^{\lambda \mu \nu} + 
K^{\mu \nu \lambda} + K^{\nu \lambda \mu}$. Note that, when identifying the basic 
{\it contravariant} tensors $z^\mu$ and $V^{\mu \nu}$, the tensor $H^{\lambda \mu \nu}$
(and the force density ${\cal F}^\mu$) must be identical in our conventions 
and in the ones of Refs.~\cite{BS95}, \cite{BS96} (who use the opposite signature).
However, our result Eq.~(\ref{eqh}) differs, 
after the substitution $G\lambda \rightarrow f_a/8$,  in many terms from the second 
Eq.~(31) of Ref.~\cite{BS96}. Whatever be the corrections we could think 
of doing on the second term in their Eq.~(31) (which is dimensionally 
wrong, probably by a copying error leading to a forgotten overdot on one of the two terms), 
we saw no way of reconciling their result with ours (even after 
expanding explicitly $V^{\mu \nu} = \dot{z}^\mu \, {z}^{\nu \prime} - 
\dot{z}^\nu \, {z}^{\mu \prime}$).

\end{document}